\documentclass[twocolumn, twocolappendix, dvipsnames]{aastex631}
\let\tablenum\relax
\usepackage{amsmath,amssymb}
\usepackage{physics} 
\usepackage{siunitx}
\AtBeginDocument{\RenewCommandCopy\qty\SI}
\RenewCommandCopy\qty\SI
\ExplSyntaxOn
\msg_redirect_name:nnn { siunitx } { physics-pkg } { none }
\ExplSyntaxOff
\usepackage[capitalise]{cleveref}

\usepackage{booktabs}
\usepackage{tabularx}
\usepackage{array}
\usepackage{multirow}
\usepackage{xcolor} 
\usepackage{colortbl} 
\usepackage{cancel}

\usepackage{lipsum}

\usepackage{xspace}
\usepackage{hyperref}

\newcommand{\eg}{e.g.,\ } 
\newcommand{\ie}{i.e.,\ } 

\newcommand\Msun{\hbox{ M$_\odot$}} 
\newcommand\Mstar{\hbox{ M$_\star$}} 

\newcommand{\stkout}[1]{\ifmmode\text{\sout{\ensuremath{#1}}}\else\sout{#1}\fi}

\DeclareUnicodeCharacter{2609}{\odot}
\DeclareUnicodeCharacter{2605}{\star}

\let\sun\odot

\newcommand*\solarmass{\si{\solarmass}}
\newcommand*\solarluminosity{\si{\solarluminosity}}
\DeclareSIUnit\lightspeed{$c$}
\DeclareSIUnit\rydberg{Ry}
\DeclareSIUnit\erg{erg}
\DeclareSIUnit\magnitude{mag}
\DeclareSIUnit\mag{mag}
\DeclareSIUnit\jansky{Jy}
\DeclareSIUnit\gauss{G}
\DeclareSIUnit\h{$h$}
\DeclareSIUnit\hseven{$h$_7}
\DeclareSIUnit\parsec{pc}
\DeclareSIUnit\year{yr}
\DeclareSIUnit\solarluminosity{\ensuremath{L_\sun}}
\DeclareSIUnit\solarmass{\ensuremath{M_\sun}}
\DeclareSIUnit\solarmassinenergy{\ensuremath{M_\sun|c^2}}
\DeclareSIUnit\solarradius{\ensuremath{R_\sun}}
\DeclareSIUnit\astronomicalunit{AU}
\DeclareSIUnit\clight{\ensuremath c}

\newcommand{\IAIFI}{The NSF AI Institute for Artificial Intelligence and Fundamental Interactions}
\newcommand{\CfA}{Center for Astrophysics $|$ Harvard \& Smithsonian, Cambridge, MA 02138, USA}
\newcommand{\MIT}{Department of Physics, Massachusetts Institute of Technology, Cambridge, MA 02139, USA}
\def \RTPVarGalTotal{196\xspace} 
\def \RTPVarCSSTotal{23,351\xspace} 
\def \RTPVarGCTotal{22,828\xspace} 
\def \RTPVarUCDTotal{523\xspace} 

\hypersetup{linkcolor=red,citecolor=blue,filecolor=cyan,urlcolor=magenta}

\usepackage{soul}
\received{December 23, 2024}
\revised{May 21, 2025}
\accepted{May 30, 2025}

\submitjournal{ApJ}

\begin{document}

\title{A Wide Field Map of Ultra-Compact Dwarfs in the Coma Cluster}

\correspondingauthor{Richard T. Pomeroy}
\email{richard.pomeroy01@utrgv.edu}

\author[0000-0003-1595-9903]{Richard T. Pomeroy}
\affiliation{Department of Physics and Astronomy, The University of Texas Rio Grande Valley, Brownsville, TX 78520, USA}

\author{Juan P. Madrid}
\affiliation{Department of Physics and Astronomy, The University of Texas Rio Grande Valley, Brownsville, TX 78520, USA}
\affiliation{National Science Foundation, 2415 Eisenhower Avenue, Alexandria, VA 22314, USA}

\author{Conor R. O'Neill}
\affiliation{Australian Astronomical Observatory, PO Box 915, North Ryde, NSW 1670, Australia}

\author[0000-0003-4906-8447]{Alexander T. Gagliano}
\affiliation{\IAIFI}
\affiliation{\CfA}
\affiliation{\MIT}







\begin{abstract}

A dataset of \RTPVarCSSTotal globular clusters (GCs) and ultra-compact dwarfs (UCDs) in the Coma cluster of galaxies was built using Hubble Space Telescope Advanced Camera for Surveys data. Based on the standard magnitude cut of $M_V\!\leq\!-11$, a total of \RTPVarUCDTotal UCD candidates are found within this dataset of Compact Stellar Systems (CSS). From a color-magnitude diagram (CMD) analysis built using this catalog, we find a clear mass-magnitude relation extending marginally into the UCD parameter space. The luminosity function defined by this dataset, shows an excess of sources at bright magnitudes, suggesting a bimodal formation scenario for UCDs. We estimate the number of UCDs with a different origin than GC to be $N_\text{UCD}\gtrsim32\pm1$. We derive the total number of CSS within the core ($\qty{1}{\mega\parsec}$) of Coma to be $N_{CSS}\!\approx\!\num[group-separator = {,}]{69400(1400)}$. The radial distribution of UCDs in Coma shows that, like GCs, UCDs agglomerate around three giant ellipticals: NGC 4874, NGC 4889, and IC 4051. We find UCDs are more centrally concentrated around these three ellipticals than GCs.
IC 4051 has a satellite population of UCDs similar to NGC 4874 and NGC 4889. We estimate only $\sim\!14\%$ of UCDs, inhabit the intracluster space (ICUCD) between galaxies in the region, in comparison to $\sim\!24\%$ for GCs (ICGC). We find red (metal-rich) UCDs are more likely located closer to a host galaxy, with blue (metal-poor) UCDs showing a greater dispersion and lower average density in the region.


\end{abstract}



\section{Introduction} \label{sec:intro}

Ultra-Compact Dwarfs (UCDs) are considered to be the missing link between globular clusters (GCs) and compact elliptical (cEs) galaxies. Specifically, by exploring the Fundamental Plane relations of hot stellar systems, \citet{Misgeld_2011} and more recently, \citet{Wang_2023} have found that in the size-magnitude plane UCDs populate a parameter space between GCs and cE galaxies. Some UCDs exhibit a similar mass-size relation to massive ellipticals, cEs, and nuclear star clusters \citep{Misgeld_2011, Norris_2014, Wang_2023}. Interestingly, other authors have found that UCDs might be the densest galaxies in the local universe \citep{Strader_2013}.

The observational properties commonly used to define UCDs are: their magnitudes ($M_V \qty{\leq -11}{\mag}$), at the bright end of the globular cluster luminosity function, and their sizes (effective radii, $R_e\qty{>10}{\parsec}$) \citep{Mieske_2006}. The term UCD has been more persistent in the literature through time than other terminology, such as Dwarf Galaxy Transition Object \citep[DGTO;][]{Hasegan_2005}, or intermediate-mass objects \citep[IMO, see][]{Hilker_2006, Kissler-Patig_2006} that refer to stellar entities of similar characteristics.

UCDs have been found across virtually all galactic environments: galaxy clusters \citep[\eg][]{Hilker_1999, Drinkwater_2000, Mieske_2004, Mieske_2007, Blakeslee_2008, Chilingarian_2008, Caso_2014}, fossil groups \citep{Madrid_2011, Madrid_2013}, Hickson compact groups \citep{DaRocha_2011}, and low-density environments \citep{Hau_2009} -- see also a compilation by \citet{Bruns_2012}. In fact, more than a decade ago \citet{Norris_2011} postulated the ubiquity of UCDs in all environments.

Research into UCDs over the last two decades \citep[\eg][]{Thomas_2008, DaRocha_2011, Pfeffer_2013, Norris_2015, Pfeffer_2016, Goodman_2018, Mahani_2021, Khoperskov_2023, Wang_2023} has suggested multiple formation pathways for these objects. UCDs could represent a continuation of the high mass end of GCs in a galaxy cluster formation scenario \citep{Mieske_2002, Bekki_2002} or alternatively the tidally stripped remnants of nucleated dwarf galaxies after an encounter with a larger galaxy remove most of the extended outer structure of the dwarf, but leave the core relatively intact \citep{Bekki_2001}. A recent study of the Virgo cluster by \citet{Wang_2023}, observed objects which fit the morphological space between nucleated dwarf galaxies and UCDs, revealing a transient evolutionary stage and further supporting the tidal threshing hypothesis. However, many authors \citep[\eg][]{Norris_2014, Pfeffer_2016, Saifollahi_2021} suggest the UCD population is a composite of the two formation methods, with overlap below a star cluster formation limit of $M\lesssim\qty{5E7}{\Msun}$ \citep{Norris_2019}. Studies of the Fornax cluster \citep{Wittmann_2016, Saifollahi_2021} and the Virgo cluster \citep{Liu_2015, Liu_2020}, have demonstrated the value of systematic analysis of UCDs in galaxy cluster environments where accretion and mergers are evident, and have shown that it can be informative to study the populations of these objects, especially as tracers of dark matter and as the fossil remnants of the turbulent evolutionary history of galaxy clusters.

An interesting aspect of the dense stellar environments in the cores of Compact Stellar Systems (CSS, \ie GCs, UCDs, and cEs) is that they may contain central intermediate mass black holes (IMBH) or even supermassive black holes (SMBH), in the case of stripped UCDs. The mass of a central black hole would therefore be a distinguishing feature between the formation pathways.  SMBHs have been detected in 5 putative stripped nucleus type UCDs \citep[see][]{Seth_2014, Ahn_2017, Ahn_2018, Afanasiev_2018}. Furthermore, \citet{Voggel_2018} have determined the upper limits for SMBH in 2 further UCDs. More recently, \citet{Pechetti_2022} suggest the presence of an IMBH \qty{\sim E5}{\Msun} in a stripped nucleus UCD around M31. Based on these results, many authors postulate the existence of a large, under-reported, population of supermassive black holes in UCDs. However, despite many and varied searches \citep[\eg see][for recent examples]{Gomez_2023, Pomeroy_2024, Tang_2024} a conclusive proof for IMBH in CSS has not yet been found.

The presence of small samples of UCDs in the Coma cluster of galaxies has been well documented. \citet{Price_2009}, using HST data obtained before the ACS failure, reported the existence of seven compact and luminous stellar systems with the characteristics of cEs and UCDs. \citet{Adami_2009} obtained spectra of five UCDs in the Coma cluster.

The focus of this work is to describe the cluster-wide distribution of UCDs in the Coma cluster (Abell 1656) using a dataset of GCs and UCDs described in the next section (§\ref{sec:data_methods}), where methods used with justification are also referenced. The CSS analysis begins with a description of the color-magnitude diagram (§\ref{sec:CMD_GCs_and_UCDs}), while the CSS luminosity function is explored in §\ref{sec:CSS_LF}. The radial profile, spatial distribution, intracluster fraction, and distribution by magnitude and color of UCDs are discussed in section §\ref{sec:WideFieldMap}, with final conclusions in §\ref{sec:Conclusions}.

A distance to Coma of \qty{100}{\mega\parsec} ($(m-M)\qty{=35.0}{\mag}$) is adopted \citep{Carter_2008}.

\section{Data and Methods} \label{sec:data_methods}

The details of the original dataset were presented in an earlier paper: \citet{Madrid_2018}. An augmented dataset used in this paper of \RTPVarCSSTotal CSS, consisting of \RTPVarGCTotal GCs and \RTPVarUCDTotal UCDs in Coma was built using 26 pointings of the Advanced Camera for Surveys (ACS). 

The dataset we use was, as detailed in \citet{Madrid_2018}, built following an onerous eye inspection of candidates to verify that the morphology was compatible with a CSS at the distance of Coma. CSS were selected based on the analysis of their magnitudes, colors, sizes and morphologies.

Here, we briefly summarize the Appendix A of \citet{Madrid_2018} that details the steps we took to build the dataset of CSS used in this study.

It is well established \citep[\eg][]{Larsen_2001}, that extragalactic CSS populate a well defined parameter space in a color magnitude diagram, with colors generally in the range $0.5\!<\!(F475W - F814W)\!<\!2.5$. By creating color cuts within the CMD and visually inspecting outliers we were able to identify background galaxies with obvious spiral arms, other background objects with elongated morphology, and steep gradients of galaxy light. Similarly, by doing an analysis of the sizes of the candidates, and displaying those objects that were too big or two small compared with the expected size of a CSS at the distance of Coma (i.e. a few pc for a GC; $\sim10$ to tens of pc for a UCD) we were able to clean background objects, cosmic rays, and other artefacts. At the distance of Coma, CSS should be either unresolved or marginally resolved by HST data. Those objects that showed photometric or morphological properties outside the parameters expected for globular clusters were scrutinized on the screen using the images in the two filters available to us.

\begin{figure}[t]
    \centering
    \includegraphics[width=1\linewidth]{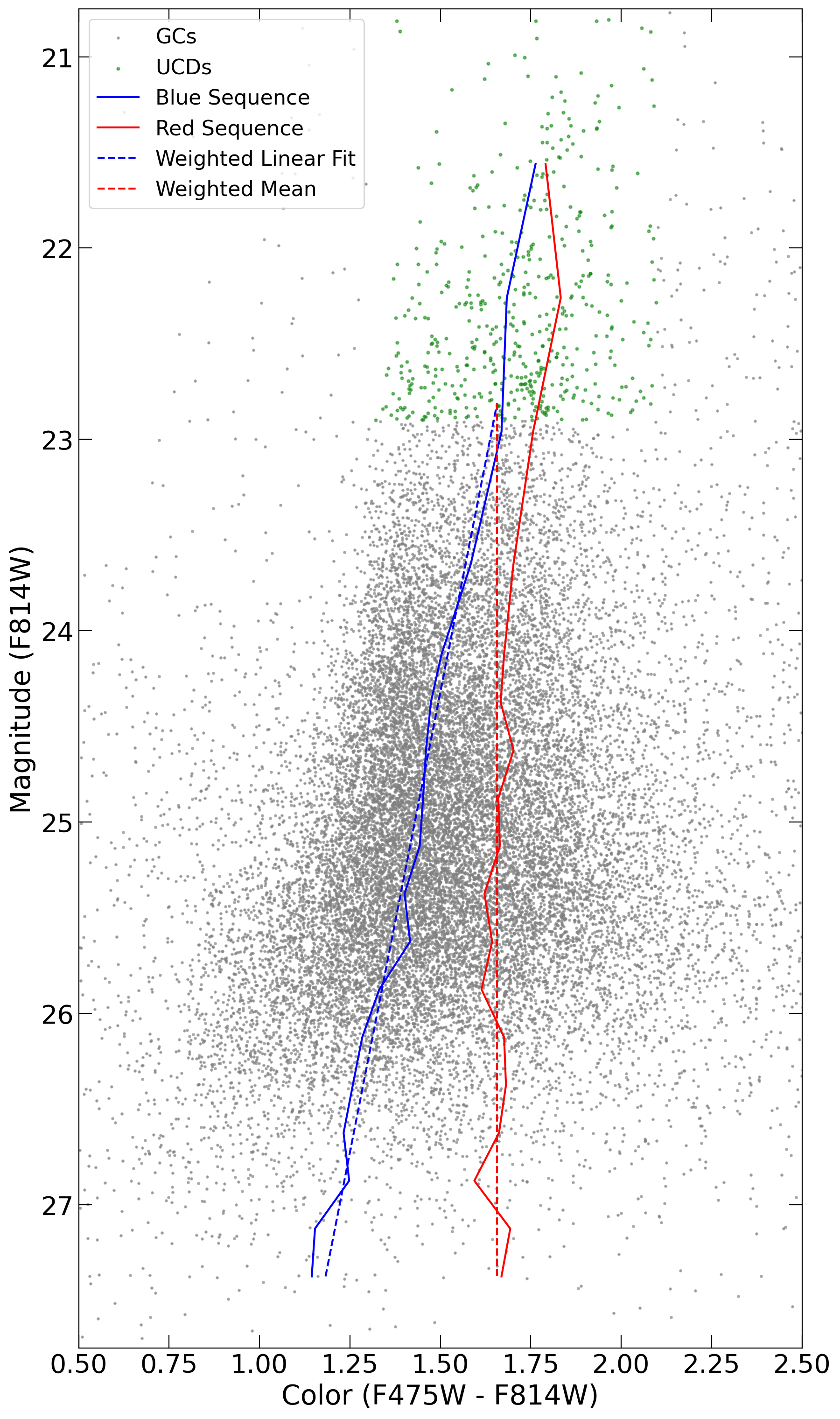}
    \caption{Color–magnitude diagram (CMD) with magnitude $(F814W)$ vs. color $(F475W\!-\!F814W)$ for the data used in the bimodal sequence fits (see the text). The solid lines connect the mean points from \cref{tab:cmd_unconstrained_fits} representing the unconstrained bimodal fits.}
    \label{fig:Coma_CSS_CMD}
\end{figure}

The augmented dataset of \RTPVarCSSTotal CSS used in this paper compares to 22,426 used in \citet{Madrid_2018} after detailed corrections and improvements were made to allow inclusion of data points which were previously rejected as spurious. These improvements involved correlation of source extracted objects with the original images in both filters, primarily close to galaxy cores to confirm the existence of real sources. This is a well known issue with \texttt{DAOPHOT} which has a propensity for identifying false positives in steep light gradients close to galaxy cores, necessitating careful manual inspection.

The details of the ACS data used for this work are also given in \citet{Madrid_2018}. The ACS pointings cover the core of Coma and its two brightest cluster galaxies (BCGs): NGC 4889, and NGC 4874. The ACS data also include IC 4051 a giant elliptical galaxy that has a large population of GCs \citep{Woodworth_2000, Madrid_2018}. The ACS data was obtained using two filters: $F475W$ (similar to Sloan $g$) and $F814W$ (similar to Cousins $I$).

Selection of UCD candidates by combining HST photometry, that is, their color and magnitude and their morphological information has been shown to work effectively. \citet{Madrid_2010} found 52 UCD candidates in a single ACS pointing containing one of the two Brightest Cluster Galaxies of Coma: NGC 4874. Five of the above 52 candidates were included in a Coma-wide spectroscopic survey carried out with the Keck telescope by \citet{Chiboucas_2011}. All of five candidates with spectroscopic data were confirmed as genuine Coma UCDs.

The efficacy of the method used here to select UCD candidates in Coma was also shown to work in the Fossil group NGC 1132 which is located at roughly the same distance as Coma ($D\qty{\sim100}{\mega\parsec}$) by \citet{Madrid_2011, Madrid_2013}. This method was also successfully applied to ACS data of the Abell cluster 1689 where 160 UCD candidates were found \citep{Mieske_2004}. More recently similar methods have been used by \citet{Harris_2020} with HST / ACS data as part of an ongoing study into the GCs in the Perseus cluster.

As mentioned in the Introduction (§\ref{sec:intro}), UCDs are generally considered to have absolute magnitudes $M_V \qty{\leq-11}{\mag}$. By assuming $(V-I)\qty{=1.1}{\mag}$ and a distance to the Coma cluster of $D\qty{\sim100}{\mega\parsec}$, or $(m-M)\qty{=35.0}{\mag}$ \citep{Carter_2008}, the apparent magnitude threshold for UCDs is $F814W \qty{<22.9}{\mag}$. For the purpose of this study, we therefore took objects with a magnitude brighter than $F814W \qty{<22.9}{\mag}$ and color between $1.3 < (F475W - F814W) < 2.1$ to be the UCD candidates. Using this criteria we found \RTPVarUCDTotal UCD candidates in our dataset.

\begin{figure}[t]
    \centering
    \includegraphics[width=1\linewidth]{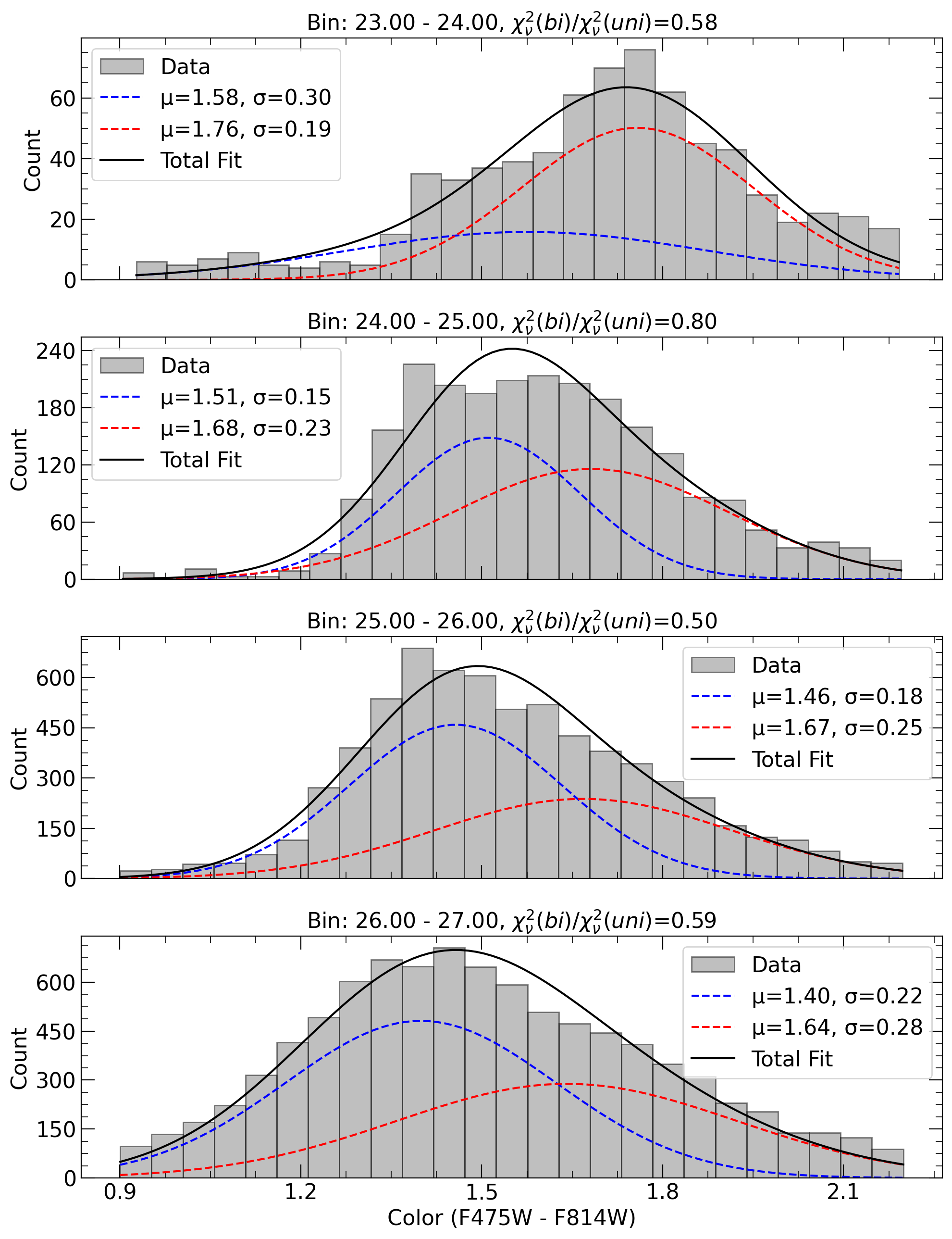}
    \caption{Sample Gaussian Mixture Model (GMM) solutions for the $(F475W - F814W)$ color distributions in four 1 mag bins in $F814W$ as labeled. In each panel, the dashed lines show the Gaussian curves matching the blue and red sequences, while the solid lines show the sum of the two components. The ratio $\chi^2_\nu(bi)/\chi^2_\nu(uni)<1$ indicates a bimodal Gaussian is a better fit to the data.}
    \label{fig:Sample_GMM}
\end{figure}

\begin{deluxetable*}{c|c|c|c|c|c|c|c}  
    \tablecaption{Bimodal fits to the $(F475W - F814W)$ color - Unconstrained \label{tab:cmd_unconstrained_fits}}
    \tablehead{
        \colhead{Magnitude Range} & \colhead{N} & \colhead{$\mu_1(\pm)$} & \colhead{$\sigma_1(\pm)$} & \colhead{$\mu_2(\pm)$} & \colhead{$\sigma_2(\pm)$} & \colhead{$p_1(\pm)$} & \colhead{$\dfrac{\chi^2_\nu(bi)}{\chi^2_\nu(uni)}$} 
    }
    \startdata
            21.20 - 21.90 & 59 & 1.763(0.049) & 0.329(0.052) & 1.791(0.184) & 0.305(0.074) & 0.21(0.13) & 1.39 \\
	21.90 - 22.60 & 96 & 1.684(0.181) & 0.344(0.064) & 1.833(0.031) & 0.076(0.063) & 0.54(0.20) & 0.54 \\
	22.60 - 23.30 & 239 & 1.686(0.028) & 0.286(0.043) & 1.744(0.035) & 0.175(0.049) & 0.58(0.09) & 0.88 \\
	23.30 - 24.00 & 692 & 1.583(0.017) & 0.200(0.022) & 1.704(0.017) & 0.211(0.019) & 0.53(0.04) & 1.27 \\
	24.00 - 24.25 & 2010 & 1.516(0.010) & 0.160(0.009) & 1.709(0.017) & 0.232(0.005) & 0.57(0.06) & 0.83 \\
	24.25 - 24.50 & 1191 & 1.474(0.017) & 0.167(0.022) & 1.667(0.018) & 0.229(0.007) & 0.56(0.07) & 1.13 \\
	24.50 - 24.75 & 1539 & 1.460(0.016) & 0.156(0.019) & 1.702(0.034) & 0.240(0.008) & 0.61(0.09) & 0.41 \\
	24.75 - 25.00 & 1821 & 1.452(0.010) & 0.183(0.009) & 1.660(0.021) & 0.243(0.005) & 0.58(0.07) & 0.47 \\
	25.00 - 25.25 & 2177 & 1.443(0.012) & 0.184(0.012) & 1.664(0.024) & 0.257(0.005) & 0.60(0.08) & 0.48 \\
	25.25 - 25.50 & 2384 & 1.401(0.020) & 0.177(0.013) & 1.622(0.030) & 0.261(0.008) & 0.41(0.08) & 0.38 \\
	25.50 - 25.75 & 2629 & 1.416(0.008) & 0.224(0.008) & 1.643(0.019) & 0.275(0.005) & 0.55(0.06) & 0.69 \\
	25.75 - 26.00 & 2332 & 1.331(0.028) & 0.199(0.017) & 1.614(0.036) & 0.283(0.017) & 0.42(0.06) & 0.56 \\
	26.00 - 26.25 & 1787 & 1.284(0.017) & 0.192(0.010) & 1.676(0.038) & 0.256(0.015) & 0.56(0.06) & 0.48 \\
	26.25 - 26.50 & 1101 & 1.258(0.019) & 0.192(0.011) & 1.681(0.052) & 0.262(0.019) & 0.59(0.08) & 0.63 \\
	26.50 - 26.75 & 616 & 1.233(0.028) & 0.194(0.014) & 1.662(0.037) & 0.274(0.017) & 0.54(0.06) & 0.69 \\
	26.75 - 27.00 & 309 & 1.248(0.079) & 0.222(0.050) & 1.594(0.040) & 0.292(0.024) & 0.52(0.07) & 1.04 \\
	27.00 - 27.25 & 141 & 1.153(0.046) & 0.154(0.033) & 1.693(0.066) & 0.256(0.021) & 0.54(0.09) & 0.54 \\
	27.25 - 27.50 & 70 & 1.144(0.146) & 0.198(0.107) & 1.669(0.083) & 0.275(0.041) & 0.45(0.14) & 0.95 \\
        \enddata
\end{deluxetable*}

\section{Colour-Magnitude diagram for GCs and UCDs} \label{sec:CMD_GCs_and_UCDs}
In \cref{fig:Coma_CSS_CMD} we present the Color Magnitude Diagram (CMD) derived from our dataset. In the CMD, objects with magnitude brighter than $F814W\!<\!\qty{22.9}{\mag}$ and color between $1.3\!<\!(F475W\!-\!F814W)\!<\!2.1$ are plotted in green; these are the \RTPVarUCDTotal UCD candidates. The purpose of this section is to identify the existence of blue and red subpopulations of CSS and to characterize these sequences as a function of magnitude and color, \ie the Mass-Metallicity Relation (MMR), \citep[see \eg][]{Strader_2006}. In order to characterize the two subpopulations of CSS in our data, we follow a process similar to that adopted by \citet{Harris_2009} for their analysis of the globular cluster system in M87.

The process involves separating the color-magnitude data into narrow magnitude bins as detailed below, based on available sample size and fitting a Gaussian Mixture Model (GMM) to the data in each of those magnitude bins. The purpose of splitting into magnitude bins is to ensure no apriori assumptions are made with regards to the form of the MMR.

\begin{figure}[t]
    \centering
    \includegraphics[width=1\linewidth]{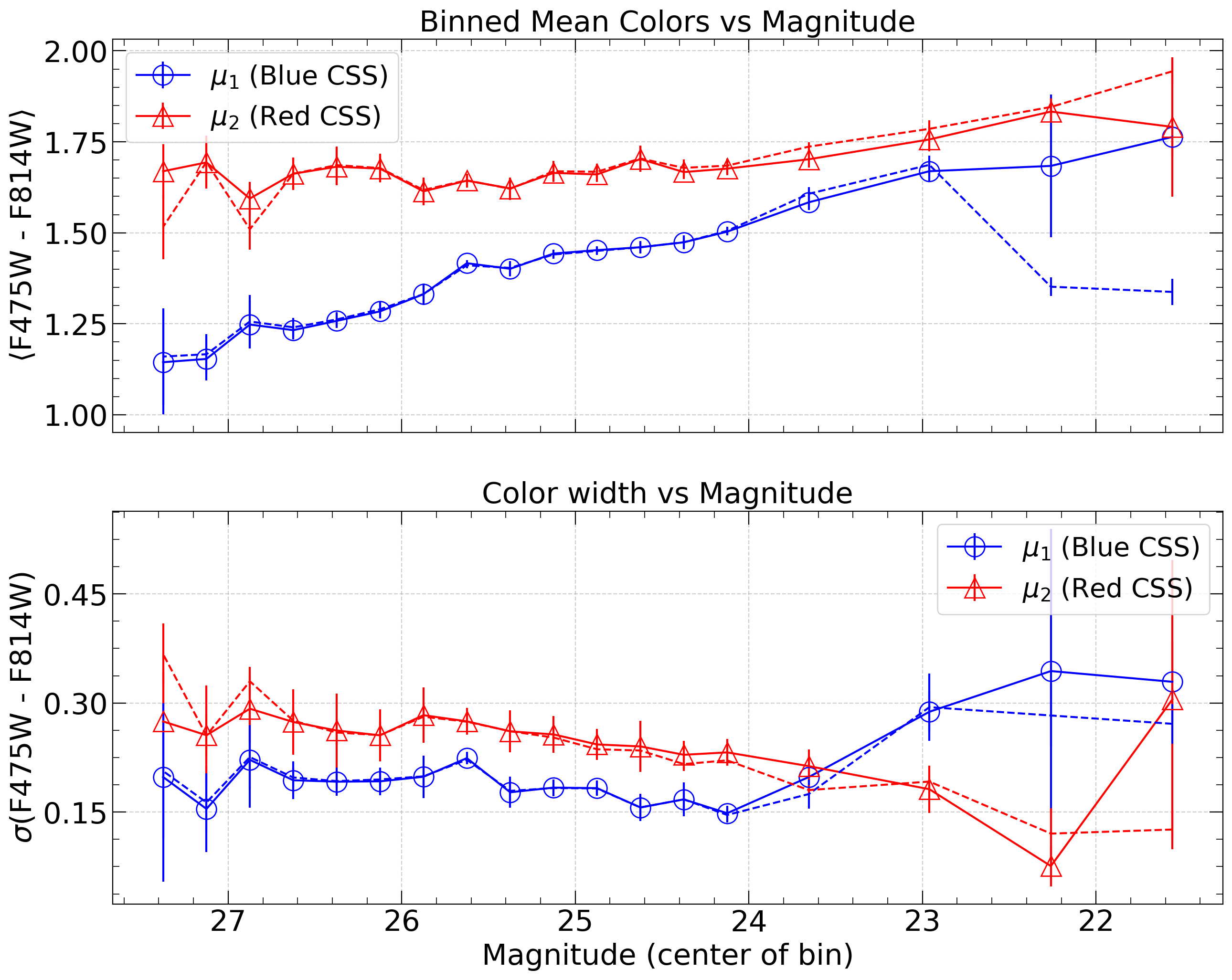}
    \caption{(Top) Binned mean colors in $(F475W\!-\!F814W)$ for the blue (open circles) and red (open triangles) sequences, as listed in \cref{tab:cmd_unconstrained_fits}. (Bottom) Internal standard deviation (color width) of the blue and red sequences as a function of $F814W$ magnitude. For the blue sequence, the dashed lines show the change to the fitted color and dispersion if the red sequence is constrained to have an initial expectation value $\mu_2 = 1.657$ at all magnitudes. Similarly, for the red sequence, the dashed lines show the change when the blue sequence initial expectation value is constrained to have color $\mu_1 = 1.417$ at all magnitudes.}
    \label{fig:meancol_colwidth}
\end{figure}

We use the python code \texttt{GaussianMixture} to perform bimodal (and uni-modal) fitting to our data. The approach taken is to allow the \texttt{GaussianMixture} function to freely solve to the data without any restriction on the parameters of two Gaussian components. The data we are analyzing is particularly large as it includes all CSS candidates for the core of the Coma Cluster. At the distance of Coma, our data does not show clearly delineated red and blue sequences of CSS in the CMD, unlike that noted by \citet{Harris_2009} prior to their M87 sequence fitting.  We nevertheless account for this by conservatively restricting the color range under analysis to $0.9\!<\!(F475W\!-\!F814W)\!<\!2.2$, effectively ensuring we were analyzing the bulk of the dataset. Example solutions to the data (albeit with wider magnitude bins than finally fitted) are shown in \cref{fig:Sample_GMM}. 

The parameters solved in the solution to the GMM fitting by \texttt{GaussianMixture} include the mean values for blue and red sequences, $\mu_1, \mu_2$, the corresponding dispersions of the blue and red sequences $\sigma_1, \sigma_2$ and the relative weights of the populations ($p_1, p_2$). Uncertainties on all these GMM fit values were estimated by bootstrapping with random choice data resampling. Additionally, the quality of the bimodal GMM solution in comparison to a single Gaussian solution is assessed by calculating the ratio of the reduced chi-square ($\chi^2_\nu$) values of both bimodal and uni-modal solutions.  

The ``drift'' of the blue sequence mean ($\mu_1$) to redder colors at brighter magnitudes (blue tilt) is clear in \cref{fig:Sample_GMM}, while the red sequence mean ($\mu_2$) remains more obviously constant around an average value over more than three magnitudes.

For the final analysis, the magnitude range for each sample bin was set to $F814W=\qty{0.25}{\mag}$, with the exception of the bright end of the data sample, where for $F814W<\qty{24.0}{\mag}$ a bin range of $F814W=\qty{0.7}{\mag}$ was required to get a statistically significant sample. The results of this run are shown in \cref{tab:cmd_unconstrained_fits}. The value of $\chi^2_\nu(bi)/\chi^2_\nu(uni)<1.0$ for the majority of the sample bins demonstrates the validity of the bimodal solution. The brightest bin has a limited sample size and errors. As expected, we see the merging of the two subpopulations (red and blue) in the UCD parameter space. 

In order to confirm the validity of the unconstrained solution, we also fitted the bimodal GMM  with constraints on either the initial expectation value for the red sequence mean, or the initial expectation value for the blue sequence mean. The initial expectation values for the constrained solutions were determined from the weighted means of the bins of the unconstrained solutions, which for the red sequence was $\mu_2 = 1.657$ and for the blue sequence was $\mu_1 = 1.417$. No constraints were placed on the solution in respect of the dispersions, $\sigma_1, \sigma_2$.

The results of the solutions to the different constraints are shown in \cref{fig:meancol_colwidth}, where the upper panel shows the means of the red and blue color sequences as a function of the magnitude and the lower panel shows the color width or dispersion of the red and blue color sequences, also as a function of magnitude. Uncertainties on the color are included for each of the sample bins. The dashed lines for each color sequence show the result of constraining the initial expectation value for the opposite sequence, \eg the red dashed lines show the result on the red sequence of constraining the initial expectation value of the blue sequence. As is evident, the effect of constraining the sequences has marginal effect on the opposite sequence solution in the magnitude range $24\!\lesssim\!F814W\!\lesssim\!26.5$, where sample sizes are higher, suggesting a stable solution in this region.

Metal-poor GCs are known to exhibit a mass-metallicity relation or ``blue tilt'' \citep{Harris_2006, Spitler_2006, Strader_2006}. 

The mass-metallicity relation becomes more prominent for more massive globular clusters and UCDs with $M\qty{>e6}{\Msun}$ \citep{Harris_2006}, where the red and blue sequences are seen to merge. For our analysis above, and as shown in \cref{fig:Coma_CSS_CMD}, the merge magnitude is $\approx\qty{22.8}{\mag}$ with the weighted mean, $\mu_2=1.66$. 

A model was developed by \citet{Bailin_2009} to explain the mass-metallicity relation as a self-enrichment process during which massive globular clusters ($M\qty{>e6}{\Msun}$) retain a significant fraction of supernova ejecta. The color trend shown by UCDs and bright globular clusters in \cref{fig:Coma_CSS_CMD} is in good agreement with the color trend of the models presented by \citet{Bailin_2009}.

\section{Luminosity Function} \label{sec:CSS_LF}
The quality of our data allows us to fit the luminosity function of the Coma cluster GCs. In the following analysis, we adopt a Gaussian globular cluster luminosity function (GCLF) with a turnover magnitude of $M_V = \qty{-7.4}{\mag}$ \citep[e.g.][]{Harris_1991, Jordan_2006, Peng_2008} and we set the peak of GCLF in our $F814W$ data at the distance of the Coma cluster. As noted above, $F814W$ approximates to Cousins $I$ band, and for color change from Cousins $I$ to $V$ we calculate $V-I\approx0.9(g-i) + 0.39$ from the transformations determined by \citet{Jordi_2006}. For consistency with UCD like systems, however, we adopt the marginally smaller offset of $V\approx I+1\pm0.1$ which aligns with previous works by authors such as \citet{Mieske_2012} and \citet{Harris_2009}. This allows us to use $M_I = M_V - 1\pm0.1 = -8.4\pm0.1$, and to define the GCLF turnover in our data to be $\mu_{F814W} \approx M_I + 35 = \qty{26.6(0.1)}{\mag}$.

\begin{figure}[!t]
    \centering
    \includegraphics[width=1\linewidth]{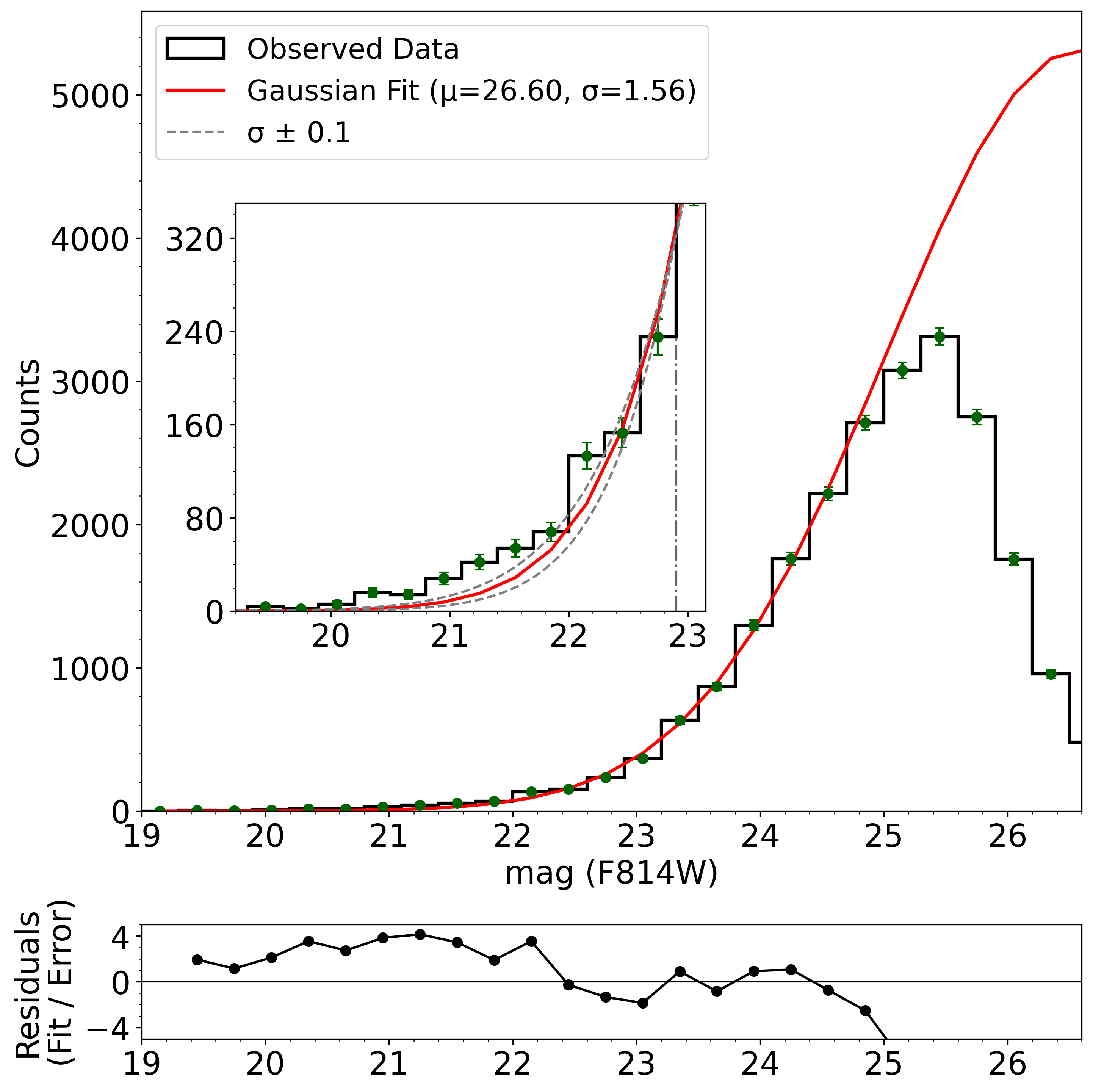}
    \caption{Luminosity function of GC and UCD candidates in our dataset. The solid red line shows the best fit of a Gaussian to the bright end of the magnitude distribution, with mean fixed at the estimated turnover of $\qty{26.6(0.1)}{\mag}$ as discussed in the text. Poisson uncertainties ($1\sigma$) are shown in green. The inset shows a zoom on the magnitude range characteristic of UCDs with the dot-dashed vertical line demarcating the \qty{22.9}{\mag} UCD limit. The dashed gray lines show $\sigma\pm0.1$ solutions, normalized to the \qty{22.9}{\mag} UCD count. The bottom panel shows the residual of the Gaussian fit to the dataset.}
    \label{fig:CSS_GCLF}
\end{figure}

We use the python package \texttt{emcee.EnsembleSampler} to fit a Gaussian to the dataset on the bright side of the distribution ($22.0\!<\!F814W\!<\!25.0$), where completeness is estimated to be acceptable (\ie $\gtrsim\!90\%$). To ensure sufficient samples in our fixed bins, we adopt a bin width of $\qty{0.3}{\mag}$ and perform 5000 samples in the Monte Carlo sampling. The results of this fitting give Gaussian parameters of $\sigma = 1.562\pm0.013$, for $\mu = 26.6\pm 0.1$.

\begin{deluxetable}{ccccc}
	\centering
	\caption{Observed UCD candidates (Obs) compared with predicted UCD count (Pred) from Gaussian modeling described in the text. Uncertainties are $1\sigma$. Masses estimated using g-band mass-to-light ratio of 3.25.}
	\label{tab:excess}
\tablehead{ 
     \colhead{Threshold} & \colhead{Mass} & \colhead{Obs} & \colhead{Pred} & \colhead{Excess} \\ 
     \colhead{(mag)} & \colhead{(\Msun)}& \colhead{(count)} & \colhead{(count($\pm$))} & \colhead{(count($\pm$))} 
 }
\startdata
	$\lesssim22.9$ & $\num{8.3e+06}$ & 523 & $\num{187.4(7.3)}$ & $\num{335.6(7.3)}$ \\ 
	$\lesssim22.7$ & $\num{1.0e+07}$ & 384 & $\num{131.6(5.8)}$ & $\num{252.4(5.8)}$ \\ 
	$\lesssim22.0$ & $\num{2.0e+07}$ & 134 & $\num{30.6(2.0)}$ & $\num{103.4(2.0)}$ \\ 
	$\lesssim21.0$ & $\num{5.0e+07}$ & 35 & $\num{3.1(0.3)}$ & $\num{31.9(0.3)}$ \\ 
\enddata
\end{deluxetable}

\begin{deluxetable*}{c|c|c|c|c|c|c|c|c|c|c}
	\caption{Projected radial distribution of GCs (\RTPVarGCTotal candidates) and UCDs (\RTPVarUCDTotal candidates) around IC 4051, NGC 4889 and NGC 4874, as shown in \cref{fig:Coma_UCDs_radial_profile} for central region of Coma cluster. Count is $N$($dist\leq d$) and uncertainties are $1\sigma$ Poisson errors, via bootstrap resampling.}
	\label{tab:CSS_radial_dist}
\tablehead{
	\colhead{d} & \multicolumn{2}{c}{IC 4051} & \multicolumn{2}{c}{NGC 4889} & \multicolumn{2}{c}{NGC 4874} & \multicolumn{2}{c}{Totals} & \multicolumn{2}{c}{(\%)}  \\
	\colhead{(kpc)} & \colhead{GC($\pm$)} & \colhead{UCD($\pm$)} & \colhead{GC($\pm$)} & \colhead{UCD($\pm$)} & \colhead{GC($\pm$)} & \colhead{UCD($\pm$)} & \colhead{GC($\pm$)} & \colhead{UCD($\pm$)} & \colhead{GC($\pm$)} & \colhead{UCD($\pm$)}
}
\startdata
  5.0  & 130(12) & 40(6) & 147(12) & 53(7) & 63(7) & 5(2) & 340(31) & 98(15) & 1(0.14) & 19(2.87) \\
 10.0  & 394(19) & 51(7) & 617(25) & 59(7) & 358(19) & 7(3) & 1369(63) & 117(17) & 6(0.28) & 22(3.25) \\
 25.0  & 1107(32) & 72(8) & 2357(51) & 76(8) & 1590(40) & 31(5) & 5054(123) & 179(21) & 22(0.54) & 34(4.02) \\
 50.0  & 1686(39) & 76(8) & 4639(67) & 125(11) & 3898(63) & 87(9) & 10223(169) & 288(28) & 45(0.74) & 55(5.35) \\
 75.0  & 2139(47) & 81(9) & 6035(73) & 150(12) & 5644(71) & 119(11) & 13818(191) & 350(32) & 61(0.84) & 67(6.12) \\
 100.0 & 2479(49) & 91(10) & 6873(83) & 158(12) & 6960(81) & 138(11) & 16312(213) & 387(33) & 71(0.93) & 74(6.31) \\
\enddata
\end{deluxetable*}

\cref{fig:CSS_GCLF} shows the luminosity function histogram of the dataset of globular clusters and UCDs in the $F814W$ filter. Uncertainties on the histogram data are $1\sigma$ Poisson errors. The inset on \cref{fig:CSS_GCLF} illustrates more detail on the bright end of the luminosity function, that is $F814W\!<\!\qty{22.9}{\mag}$, the characteristic magnitude range of UCDs (delineated by the vertical gray dot-dashed line). The additional gray dashed lines in the inset of \cref{fig:CSS_GCLF} show the fitted $\sigma\pm0.1$, illustrating the insensitivity of the excess to changes in $\sigma$ for a fixed turnover ($\mu=\qty{26.6}{\mag}$), assuming the count at the UCD cutoff, $F814W\!<\!\qty{22.9}{\mag}$, is also fixed. The bottom panel of \cref{fig:CSS_GCLF} shows the residual, that is the difference between the Gaussian  fit to the luminosity function and the histogram itself, which have been scaled by the Poisson uncertainty. When the residual is close to zero it indicates that the Gaussian distribution is a good approximation to the data of the histogram.

The Gaussian model at a robustly defined and fixed $\mu=\qty{26.6}{\mag}$, provides a good fit to the CSS data over more than $\qty{\sim2.5}{\mag}$, from $\qtyrange{\sim22.5}{\sim25.0}{\mag}$, and significantly, this is where the majority of the CSS are observed and completeness is estimated to be above 0.9. The Gaussian fit, naturally, ceases to work on the faint end of the luminosity function due to incompleteness. More interesting, the goodness of fit of the Gaussian model declines for those bright magnitudes populated by UCDs $F814W \qty{< 22.9}{\mag}$. From \cref{fig:CSS_GCLF} it is evident that the luminosity function in the UCD range of this dataset ($M_V \qty{\leq -11}{\mag}$) is not properly fit by the simple extrapolation of the GCLF to bright magnitudes. 

To determine the excess over the Gaussian model fitted to the bright side of our data, we integrate the area under the curve to find the predicted numbers, and subtract this from the observed data in the same range. These results are shown in \cref{tab:excess}. Also included in \cref{tab:excess} are the UCD excesses predicted below magnitude thresholds of $F814W \qty{<22}{\mag}$ and $F814W \qty{<21}{\mag}$. Assuming a g-band mass-to-light ratio of 3.25 from \citet{Maraston_1998, Maraston_2005}, in respect of an old (10 Gyr), metal-poor ([Z/H] = -1.35) population based on a Salpeter IMF, the brighter of these magnitudes (\ie $F814W \approx\qty{21}{\mag}\equiv M_\text{F814W}\approx\qty{-14}{\mag}$) correlates with the star cluster formation limit of $M_\star\lesssim\!5\times10^7\Msun$ \citep{Norris_2019}. This is suggestive that there is a population of $N_\text{UCD}\!\gtrsim\!32\!\pm\!1$ UCDs in the Coma cluster which have formed through a process distinct from GCs, \ie they are not merely massive GCs. As noted in \cref{tab:excess}, taking the excess at $F814W \qty{<22.7}{\mag}$ as an equivalent threshold for a UCD cluster mass of $M_\star\gtrsim\!10^7\Msun$, we estimate that $\num{\approx252(6)}$ out of 384 UCDs, or 66\%, have formed through a process distinct from GCs. This is compatible with the findings of \citet{Pfeffer_2016} who suggested stripped nuclei UCDs account for 40\% of the GC / UCD total above $10^7 \Msun$.

Finally, integrating the Gaussian over the range of our model parameters determined for the GCLF, we find a total predicted count for CSS of $N_{CSS}\!\approx\!\num[group-separator = {,}]{69400(1400)}$.

\begin{figure}[ht]
    \centering
    \includegraphics[width=0.9\linewidth]{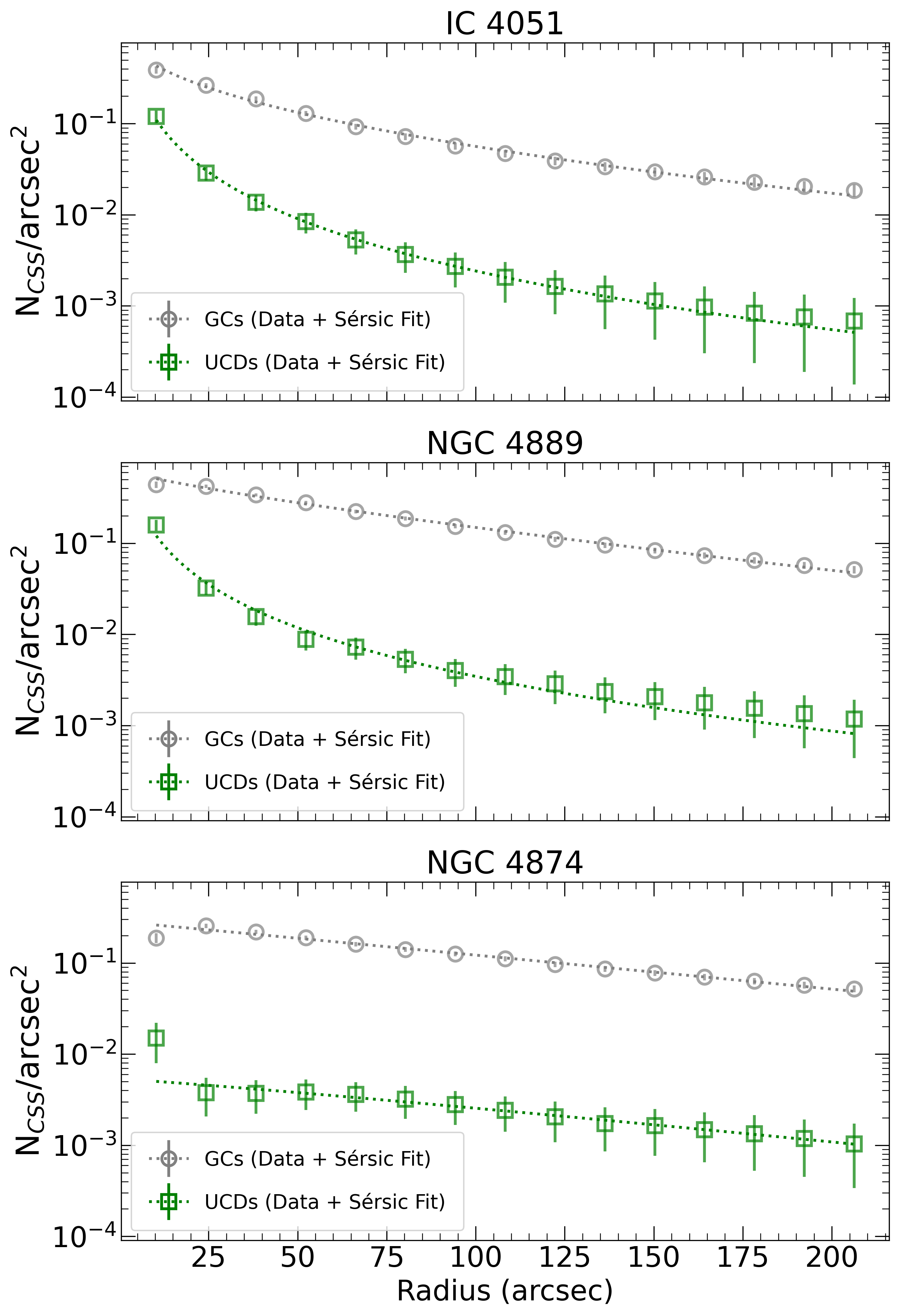}
    \caption{Radial density profiles of UCDs (green squares) in comparison to GCs (grey circles) associated with the three main overdensities around IC 4051, NGC 4889 and NGC 4874. $1\sigma$ Poisson errors are included along with fitted S\'{e}rsic profiles, the parameters for which are shown in \cref{tab:Radial_CSS_sersic}. The steeper initial reduction in UCD density with radial distance is evident, as is a clear flattening of the density about NGC 4874, which more closely follows the slope of GCs outside 50 arcseconds. We note, however, that the GC radial density profile is more likely to be affected by incompleteness close to the galaxies.}
    \label{fig:Coma_UCDs_radial_profile}
\end{figure}

\begin{deluxetable*}{c|c|c|c|c|c|c|l}
    \centering
    \caption{Parameters for S\'{e}rsic fit to radial profile of GCs and UCDs around IC 4051, NGC 4889 and NGC 4874, as shown in \cref{fig:Coma_UCDs_radial_profile} for central region of Coma cluster. Units of $R_e$ are arcseconds.}
    \label{tab:Radial_CSS_sersic}
\tablehead{
    \colhead{} & \multicolumn{2}{c}{IC 4051} & \multicolumn{2}{c}{NGC 4889} & \multicolumn{2}{c}{NGC 4874} \\
    \colhead{} & \colhead{GC} & \colhead{UCD} & \colhead{GC} & \colhead{UCD} & \colhead{GC} & \colhead{UCD} & \colhead{Description}
}
\startdata
    $n$ & 2.16 & 6.00 & 1.26 & 6.00 & 0.99 & 0.90 & S\'{e}rsic index\\
    $R_e$ & 169.09 & 80.17 & 163.91 & 122.72 & 192.94 & 186.51 & S\'{e}rsic effective radius \\
    $\Sigma_e$ & 0.0238 & 0.0037 & 0.0734 & 0.0023 & 0.0548 & 0.0012 & Surface density at $R_e$
\enddata
\end{deluxetable*} 

\section{Ultra-Compact Dwarfs in Coma} 
\label{sec:WideFieldMap}
\subsection{Radial distribution of UCDs around NGC 4874, NGC 4889, and IC 4051.}\label{sec:Radial_dist}
We count the total number of GCs and UCDs within a projected radial distance of one of the three giant ellipticals, NGC 4874, NGC 4889, and IC 4051. These results are shown in \cref{tab:CSS_radial_dist} for a sample of increasing distances out to \qty{100}{\kilo\parsec}. All distances given in this paragraph are projected distances. 

We also fit a \citet{Sersic_1968} model to these radial distribution samples, taking the radial distribution for both GCs and UCDs and fitting using 
\begin{align*}
    \Sigma(r)&=\Sigma_e\ \text{exp}\left[-b_n\left(\left(\frac{r}{r_e}\right)^{1/n}-1\right)\right]
    \intertext{with}
    b_n &= 2n-\tfrac{1}{3}
\end{align*}
and a non-linear least squares residual.

\cref{fig:Coma_UCDs_radial_profile} shows a direct comparison of UCD and GC radial density profiles, with fitted S\'{e}rsic profiles, the parameters for which are given in \cref{tab:Radial_CSS_sersic}. The uncertainties in the profile are $1\sigma$ Poisson errors, determined through bootstrap resampling. Given a total count of UCD candidates in our sample of \RTPVarUCDTotal, we determine \qty{34\pm4.4}{\%} of UCDs are found within \qty{25}{\kilo\parsec} of one of these three ellipticals, compared with only \qty{22\pm0.5}{\%} of the total count of \RTPVarGCTotal GCs within the same radius. Furthermore, this trend continues to much greater radial distances, with \qty{74\pm6.1}{\%} of all UCDs found within \qty{100}{\kilo\parsec} of one of these three ellipticals compared to \qty{71\pm0.9}{\%} of the GCs. UCDs are more likely than GCs to be found closer to one of the three giant ellipticals. This trend does not change out to \qty{100}{\kilo\parsec}.

\begin{figure*}[t]
    \centering
    \includegraphics[width=.96\linewidth]{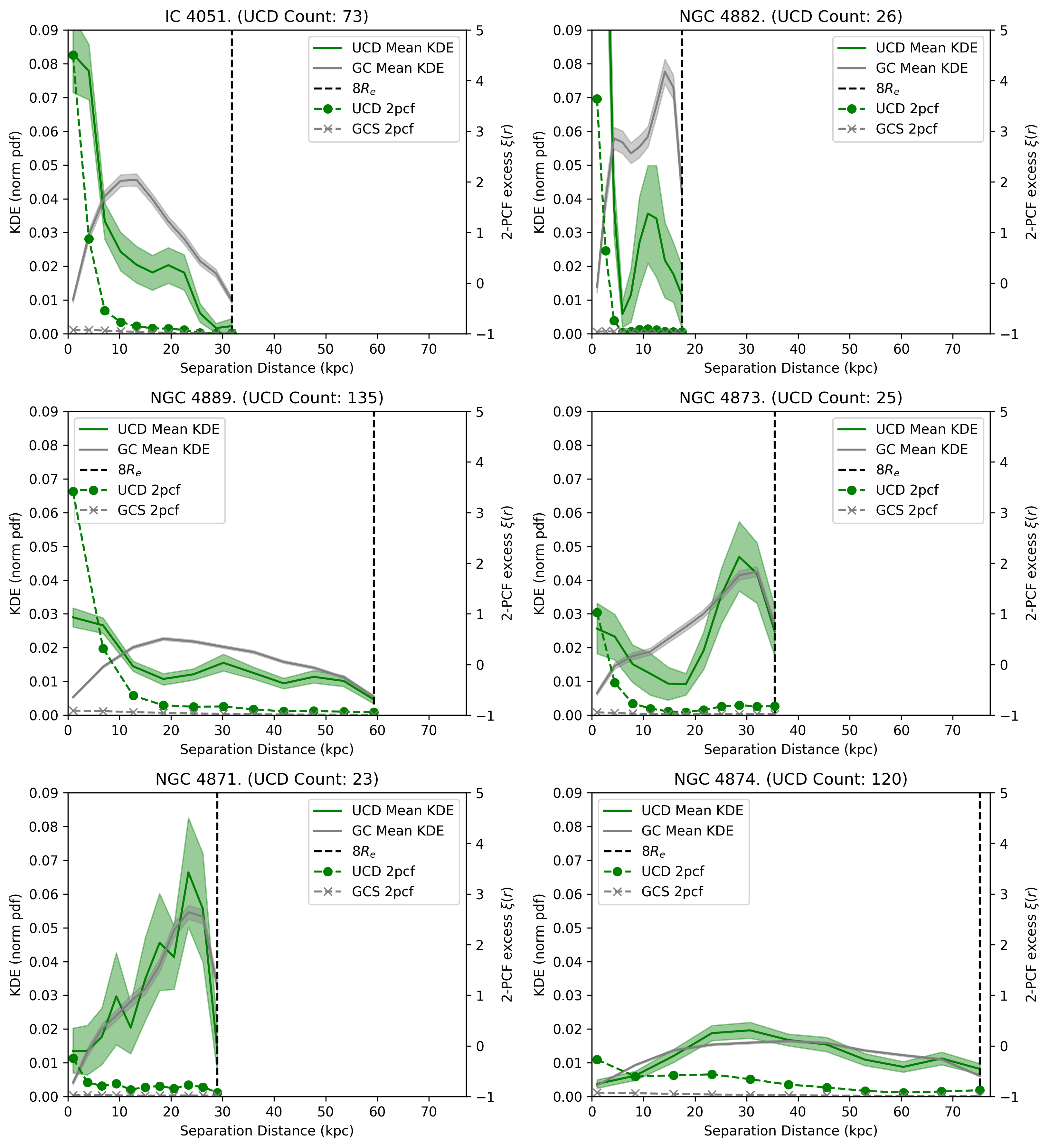}
    \caption{Kernel density estimate (KDE normalized PDF) on y1-axis and projected 2-point correlation function (2-PCF) on y2-axis, showing the excess probability of CSS clustering compared to a random distribution. Galaxies within the central region of the Coma cluster hosting $\geq15$ UCDs are included, with the radial extent being taken as $8R_e$ as illustrated in \cref{fig:Coma_UCDs_location}. The scales on the plots are the same for comparative purposes. The 2-PCF was calculated as described in the text. KDE confidence intervals were determined using bootstrap resampling. The central grouping of UCDs in both IC 4051 and NGC 4889 is evident and comparable. Conversely, while NGC 4874 has a similar order of UCDs, there is no excess compared to a random distribution. We note that NGC 4882, a satellite northwest of BCG NGC 4889, shows a high central density and excess grouping probability, but for only 26 UCDs. While we note GC data are likely incomplete close to a host galaxy, a lack of central GCs is evident in all plots, suggesting destruction of lower mass GCs compared to higher mass UCDs in CSS populations \citep[see \eg][]{Bica_2006,Madrid_2012,Madrid_2017}.}
    \label{fig:Coma_UCDs_Location_2pcf_Grid}
\end{figure*}
To quantify the clustering of UCDs/CSSs we produce a kernel density estimate (KDE) of the CSS candidates around the brightest galaxies \citep[\ie galaxies of the NGC or IC catalog suggested in][Table 1]{Madrid_2018}. With this projected radial KDE, we include a 2-point correlation function (2-PCF), to show the excess probability of a clustering of CSS compared to a random distribution. This KDE and 2-PCF are shown in \cref{fig:Coma_UCDs_Location_2pcf_Grid} for the top six galaxy hosts with $\geq15$ UCDs within $8R_e$, sorted from top-left by decreasing 2-PCF excess. The plot shows the count of UCDs within that $8R_e$ threshold. All plots have the same scale for comparative purposes.

We determine the excess probability of clustering $\xi(r)$ with radial distance from a galaxy center using
\begin{align*}
    \xi(r) &= \frac{D}{R}-1
\end{align*}

where $D$ are the counts, in separation bins, of the CSS from the galaxy center and $R$ are the counts of the separation of a random distribution from the galaxy. KDE smoothing, with a bandwidth parameter of 0.3, was applied to both the $D$ and $R$ samples to reduce sensitivity to bin-size selection. Confidence intervals on the $D$ sample were determined through 1000 bootstrap resamples. Additionally, the average $R$ value was determined through 1000 Monte Carlo simulations.

As observed above, the clustering of UCDs about the center of IC 4051 is comparable to that central to the binary BCG NGC 4889 in terms of excess. The other BCG, NGC 4874, shows no such excess compared to a random distribution, correlating with the observed wider dispersion of UCDs about this galaxy. We also note that there are at least two other galaxies, NGC 4882 (northwest of NGC 4889) and NGC 4873 (northwest of NGC 4874) with positive excess values indicative of UCD groupings, although both these galaxies have UCD counts approaching our minimum threshold.

From \cref{tab:CSS_radial_dist,fig:Coma_UCDs_radial_profile,fig:Coma_UCDs_Location_2pcf_Grid}, we can make a number of observations about the populations of CSS in Coma around the three giant ellipticals we have been studying. 
\begin{itemize}
    \item Only IC 4051, NGC 4874 and NGC 4889 host a population of UCDs in excess of 10\% of the total Coma cluster UCD candidates within $8R_e$ of the nominal galaxy center, \ie $N_\text{UCD}>52$ within $8R_e$ (see \cref{fig:Coma_UCDs_Location_2pcf_Grid}). This we consider `notable'.
    \item The projected radial density of satellite UCDs within \qty{\sim100}{\kilo\parsec} (\qty{\sim200}{\arcsec}) of IC 4051, is similar to those of the binary BCGs, NGC 4874 and NGC 4889.
    \item Out to a projected radius of \qty{\sim25}{\kilo\parsec} (\qty{\sim50}{\arcsec}) IC 4051 has a comparable number of UCDs ($72\pm8$) to the BCG NGC 4889 ($76\pm9$). This similarity is of note because IC 4051 has less than half the fractional population of GCs ($1107\pm32$) within the same projected radius compared to NGC 4889 ($2357\pm47$).
    \item A comparison between IC 4051 and the BCG NGC 4874 reveals they have similar numbers of UCDs in a projected radius of \qty{\sim50}{\kilo\parsec} (\qty{\sim100}{\arcsec}) from their centers: IC 4051 has $76\pm8$ UCDs and NGC 4874 has $87\pm9$ within that radius. Despite this similarity, however, we again note the fraction of IC 4051 UCD population is in contrast to its GC population, with IC 4051 having less than half the quantity of GCs ($1686\pm41$) in comparison to NGC 4874 ($3898\pm62$).
    \item A lack of central GCs is evident in all 6 panels of \cref{fig:Coma_UCDs_Location_2pcf_Grid}. This is suggestive of the destruction of lower mass GCs compared to higher mass UCDs in CSS populations close to their galaxy hosts, as noted by \citet{Bica_2006,Madrid_2012,Madrid_2017}.
\end{itemize}

\begin{figure*}[ht]
    \centering
    \includegraphics[width=1\linewidth]{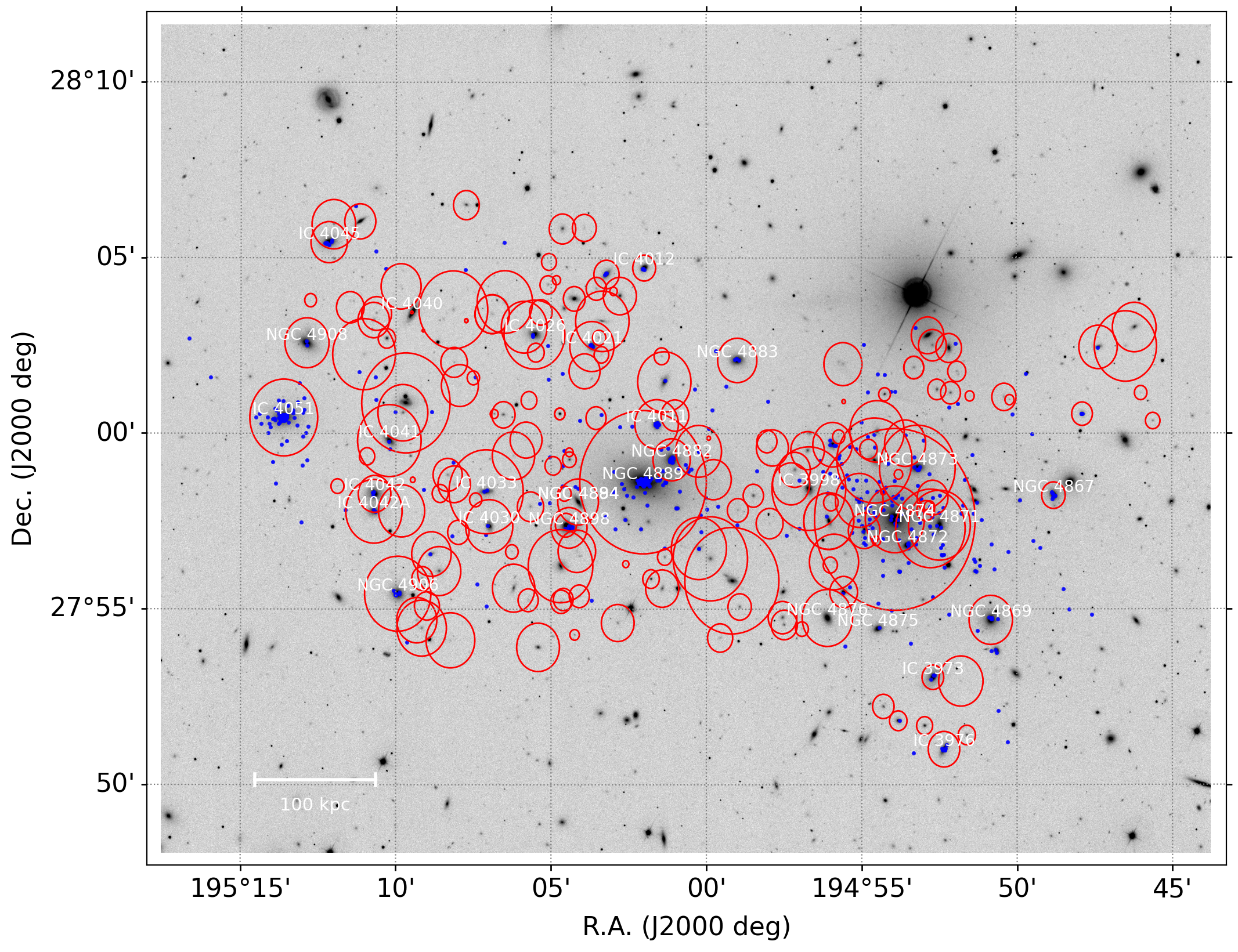}
    \caption{Location of the UCD candidates (blue dots) in the core of Coma. 
    The radial extent of galaxies are shown with red circles to $8R_e$. Three main UCD overdensities are evident around the locations of the main cluster galaxies (IC 4051, NGC 4889 and NGC 4874). While there are some UCD candidates associated with other cluster galaxies, there are also a significant number of UCDs which are outside the $8R_e$ threshold and which would be considered intracluster. The background is an SDSS g-band image. North is up and East is left.}
    \label{fig:Coma_UCDs_location}
\end{figure*}

The large concentration of UCDs around IC 4051 is even more significant in light of the fact that no other conspicuous patterns are defined by UCDs in Coma beyond their concentration around the three giant ellipticals NGC 4874, NGC 4889, and IC 4051. Of the tens of large elliptical galaxies present in the core of Coma, only the three galaxies above show a notable agglomeration of UCDs in their surroundings (\ie $N_\text{UCD}>10\%$ of the Total UCDs within $8R_e$). These agglomerations are illustrated in \cref{fig:Coma_UCDs_location} and discussed in sections §\ref{sec:Spatial_dist} \& §\ref{sec:ICUCDs}.

\subsection{Spatial distribution of UCDs in Coma}\label{sec:Spatial_dist}

\cref{fig:Coma_UCDs_location} shows a wide-field map of the UCDs in comparison to cluster member galaxies. The values for the effective radius ($R_e$) of the cluster members were taken from different sources. In \citet{Madrid_2018} the effective radius, $R_e$, was derived for NGC 4874, NGC 4889, and IC 4051 by fitting a S\'{e}rsic model \citep{Sersic_1968}. The effective radii for these galaxies can be used as a characteristic radius. Other characteristic radii can also be derived, but for this work we adopt the effective radius as an estimate of the spatial extent of these galaxies. The values of these effective radii are, for NGC 4874: \ang{;;19.4}, NGC 4889: \ang{;;15.3} for IC 4051: \ang{;;8.2}. 

Morphological data for other cluster members was taken from GalFit parameters determined by \citet{Hoyos_2011} from the original HST/ACS Coma Cluster Survey by using single S{\'e}rsic fits. Only galaxies with $R_{e}\qty{>0.5}{\arcsec}$ (spatial projection \qty{\sim 0.25}{\kilo\parsec}) were included in this analysis. Although there is some scatter in the $N_{CSS}:R_{e,gal}$ relationship, CSS populations for galaxies below this size are not significant \citep{Harris_2013}. Cluster membership was conservatively estimated based on galaxies with redshifts in the range $0.015<z<0.032$ (correlating to recession velocities $\qty{4500}{\kilo\meter\per\second}\lesssim v_r\lesssim\qty{9600}{\kilo\meter\per\second}$), to remove catalog outliers. Redshift data was taken from the eyeball catalog of Trentham et al. \citep[personal communication, and][]{Marinova_2012, Weinzirl_2014}, with which objects were correlated. This catalog provides visually determined cluster membership status for galaxies with an apparent magnitude $F814W \qty{\leq24}{\mag}$ to give a final total of \RTPVarGalTotal host galaxies. Additional effective radius parameters for IC 4040, NGC 4867, NGC 4869 and NGC 4883 were related to SDSS-r deVaucouleurs radius sourced from SDSS-DR6 \citep{Adelman-McCarthy_2008} via the \citet{https://doi.org/10.26132/ned1}, which were deemed consistent with other similar galaxies based on type and photometry and distance.

The extent of NGC 4874, NGC 4889 and IC 4051, as well as the cluster members discussed above, is taken to be $8R_e$ and indicated by the red circles in \cref{fig:Coma_UCDs_location}. 

From \cref{fig:Coma_UCDs_location}, we visually confirm the agglomerations of UCDs about IC 4051, NGC 4874 and NGC 4889, as discussed in §\ref{sec:Radial_dist}. However, we also note the presence of UCDs close to the central regions of other galaxies, for example IC 4045, IC 4042A and NGC 4908. Nevertheless, as noted in §\ref{sec:Radial_dist}, these are not significant agglomerations. The total UCD hosted populations for these galaxies being $N_\text{UCD}<15$ within $8R_e$, compared with $N_\text{UCD}>52$ within $8R_e$ for NGC 4874, NGC4889, and IC 4051.

\subsection{Intracluster UCDs}\label{sec:ICUCDs}
As can be seen in \cref{fig:Coma_UCDs_location} and as noted in §\ref{sec:Spatial_dist} above, a significant fraction of UCDs are outside $8R_e$ for galaxies in the core of the Coma cluster. If we consider this $8R_e$ limit as an `intracluster' threshold, we can estimate a count of UCDs which inhabit the intracluster space between galaxies. It has been shown that globular clusters are affected by changing potentials during accretion and merger events, and as a result migrate into the intracluster space in large numbers. \citep[\eg][]{Alamo-Martinez_2017, Lee_2022}. These intracluster globular clusters (ICGCs) are luminous tracers that are thought to provide evidence of the history of mergers within the cluster \citep[\eg][]{Harris_2020}. Additionally, authors such as \citet{Doppel_2021, Reina-Campos_2023, Lim_2024a} have shown that these ICGCs can also be used to trace the dark matter distribution in halos. Analysis of Coma ICGCs was carried out by \citet{Peng_2011} but their data did not include many additional HST pointings, significantly omitting the population around NGC 4889. Given their lower numbers, few studies have concentrated solely on intracluster UCDs, but conversely, the rich field in the Coma cluster gives us the opportunity for analysis of the UCD intracluster population compared to that of the GCs. 

Although a fixed multiple of effective radius ($R_e$) has been used to illustrate the extent of cluster galaxies in \cref{fig:Coma_UCDs_location}, it is informative to observe the fractions of both UCDs and GCs outside multiples of the effective radius for host galaxies. We therefore determine the separation of each GC and UCD to its closest galaxy. A cumulative count of objects interior to each multiple of $R_e$ is made as shown in Table \ref{tab:IC_CSS_counts}. This table includes $3\sigma$ uncertainty on the counts and the fraction of the total candidate objects (\ie GC or UCD) which this represents.

\begin{deluxetable}{c|ccc|ccc}
\centering
\label{tab:IC_CSS_counts}
\tablecaption{Cumulative count and fraction of GCs (total \RTPVarGCTotal) and UCDs (total \RTPVarUCDTotal) against multiples of galaxy effective radius, $R/R_e$, as determined in the text. The fraction external to $R/R_e$ (Ext) is also given. The figures represent cluster averages for the 196 galaxy hosts. Uncertainties are $3\sigma$. The authors reiterate the $8R_e$ threshold they consider for `intracluster' objects, as highlighted below.}
\tablehead{
    \multirow{2}{*}{$\dfrac{R}{R_e}$} & \multicolumn{3}{c}{GCs} & \multicolumn{3}{c}{UCDs} \\
    & \colhead{Count($\pm$)} & \colhead{Frac} & \colhead{Ext} & \colhead{Count($\pm$)} & \colhead{Frac} & \colhead{Ext} 
}
\startdata
     1 &  1360(3.40) & 0.06 &  0.94 & 216.81(1.06) & 0.41 & 0.59 \\
     2 &  4394(5.75) & 0.19 &  0.81 & 270.68(1.09) & 0.52 & 0.48 \\
     3 &  7670(6.76) & 0.34 &  0.66 & 327.81(1.07) & 0.63 & 0.37 \\
     4 & 10546(6.90) & 0.46 &  0.54 & 350.39(1.00) & 0.67 & 0.33 \\
     5 & 12853(6.87) & 0.56 &  0.44 & 388.29(0.92) & 0.74 & 0.26 \\
     6 & 14693(6.94) & 0.64 &  0.36 & 420.65(0.87) & 0.80 & 0.20 \\
     7 & 16144(6.74) & 0.71 &  0.29 & 429.72(0.82) & 0.82 & 0.18 \\
     \arrayrulecolor{black!30}\hline
     8 & 17351(6.00) & 0.76 &  0.24 & 447.86(0.75) & 0.86 & 0.14 \\
     \arrayrulecolor{black!30}\hline
     \arrayrulecolor{black} 
     9 & 18403(5.79) & 0.81 &  0.19 & 462.32(0.70) & 0.88 & 0.12 \\
    10 & 19263(5.35) & 0.84 &  0.16 & 476.17(0.64) & 0.91 & 0.09 
\enddata
\end{deluxetable}

We fit a S\'{e}rsic model to the radial profiles of both GCs and UCDs. The parameters for the fit are estimated as shown in Table \ref{tab:ICCSS_sersic}. This S\'{e}rsic fitting is a similar process to that carried out by \citet{Lim_2024b} in their analysis of spatial distribution of GC populations around galaxies in the Virgo cluster.

\begin{figure}[h]
    \centering
    \includegraphics[width=1\linewidth]{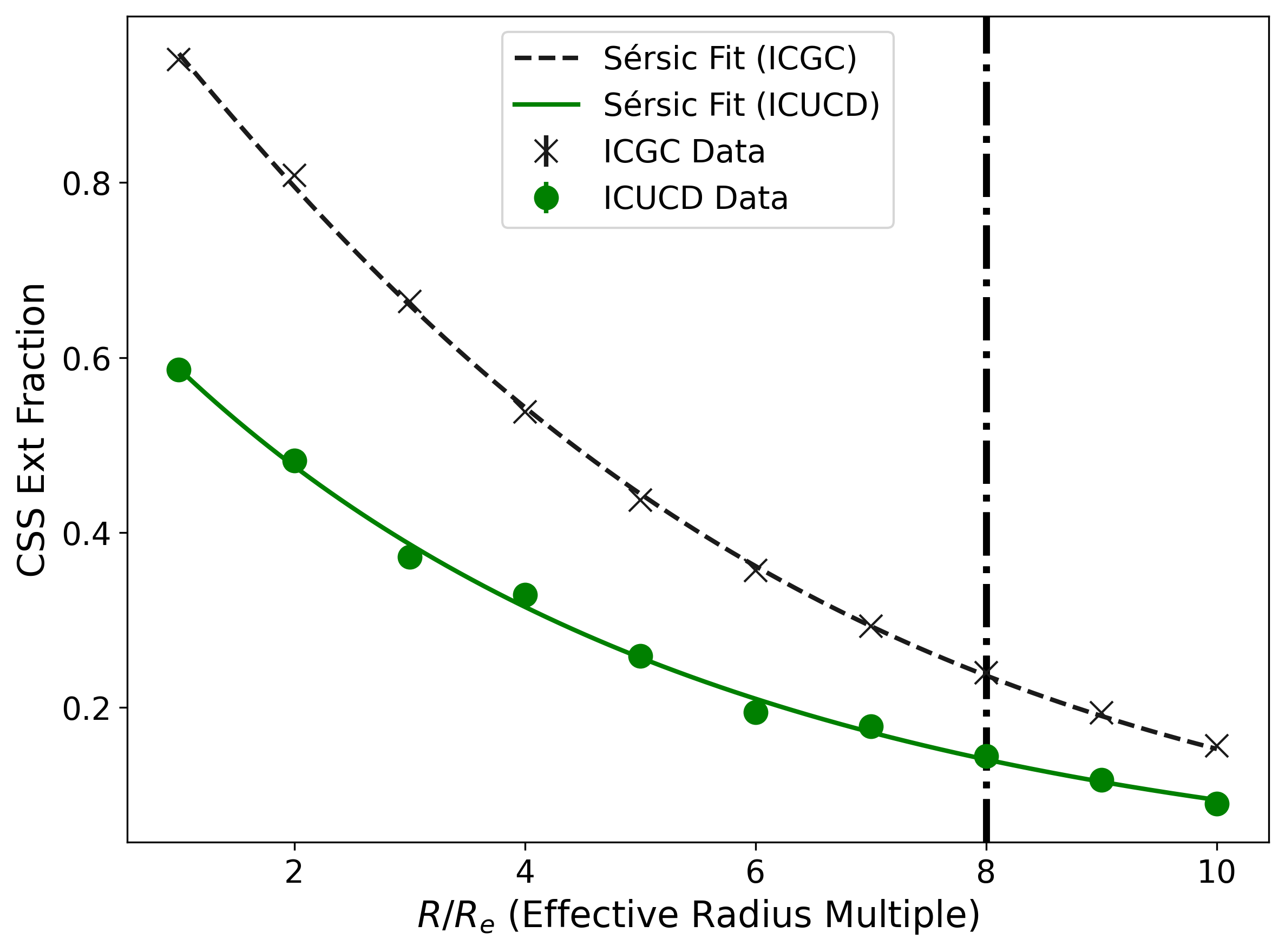}
    \caption{Comparison counts of GCs (grey dashed) and UCDs (green solid) against multiples of galaxy effective radius, $R_e$. The vertical line indicates $8R_e$ which relates to the wide-field map of UCDs (see \cref{fig:Coma_UCDs_location}) and is the threshold outside which we consider CSS to be `intracluster'. S\'{e}rsic profiles are fit to the data as described in the text. Note these radial profiles are the result of averaging over the \RTPVarGalTotal host galaxies.}
    \label{fig:ICCSS_comparison}
\end{figure}

The comparison in \cref{fig:ICCSS_comparison}, which includes these S\'{e}rsic fits, shows the percentage of CSS outside the varying multiples of effective radii where the vertical line indicates the $8R_e$ reference point used in \cref{fig:Coma_UCDs_location}. As mentioned above, taking this $8R_e$ limit to be an intracluster threshold, we find that only 14\% of UCDs are outside this hosting threshold, compared to 24\% of GCs, averaged over the 196 host galaxies. As can be determined from these data and as shown in \cref{fig:ICCSS_comparison}, the UCD population show a significantly higher relative probability of being located closer to a host galaxy. For example, $94\%$ of the GC population is exterior to $1\times R_e$ compared to $62\%$ for UCDs. Given the predicted larger mass of UCDs this difference is to be expected, and is highly suggestive that the UCDs both form closer to their host galaxies, and that due to their higher masses UCDs are less affected by the changing tidal potentials.

\begin{deluxetable}{cccc}
\label{tab:ICCSS_sersic}
\tablecaption{Best parameters for S'{e}rsic fit of GCs and UCDs shown in \cref{fig:ICCSS_comparison} for central region of Coma cluster. Units of $R_e$ are in multiples of host galaxy effective radius.}
\tablehead{
	\colhead{Param} & \colhead{GC Value} & \colhead{UCD Value} & \colhead{Description}
}
\startdata
	$n$ & 0.89 & 1.02 & S\'{e}rsic index \\
	$R_e$ & 7.58 & 8.32 & S\'{e}rsic effective radius \\
	$\Sigma_e$ & 0.26 & 0.13 & Surface density at $R_e$ \\
\enddata
\end{deluxetable}

Although the GC fraction is consistently higher than the UCD fraction, as would be expected from the radial distribution of UCDs, the ratio of GC to UCD is consistently $\sim\!1.75$, with the common profiles suggesting a common spatial distribution. This is confirmed through a Kolmogorov-Smirnov (KS) test $p-$value of 0.79. Although this is at odds with the finding of the GCLF (see §\ref{sec:CSS_LF}) we note that the GCLF is an intrinsic property of CSS, while the average cluster radial profile we have shown here is a result of the mergers and interactions of the galaxies during cluster evolution.

\begin{figure*}[ht]
    \centering
    \includegraphics[width=1\linewidth]{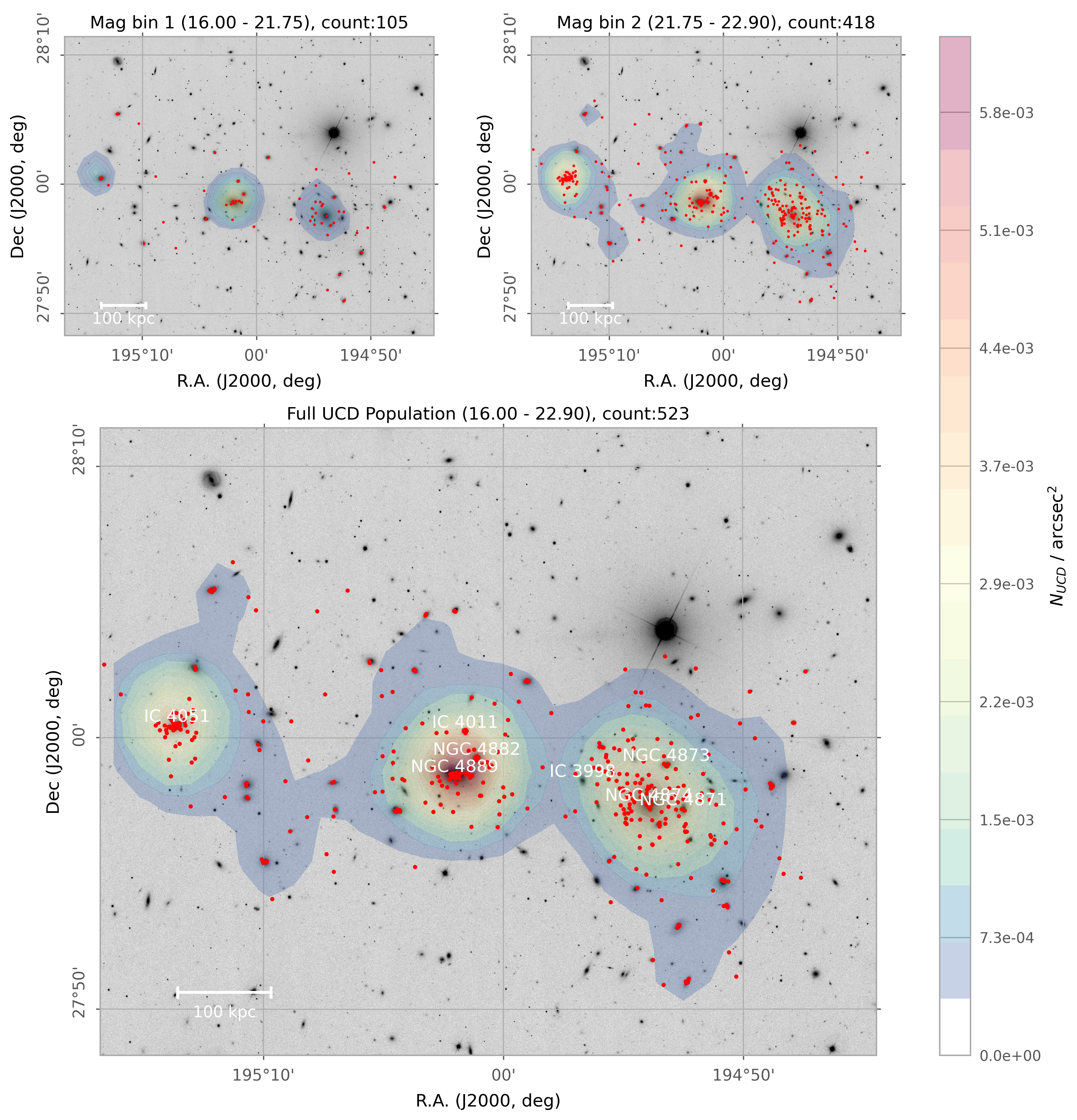}
    \caption{Spatial distribution of UCD candidates in our data, overlaid on a Coma cluster image, using magnitude splits from F814W data. Density contours are included. Top-left plot shows brightest $\qty{\sim20}{\percent}$ of UCD candidate population ($16.0<F814W<21.75$ mag) and top-right plot shows remaining $\qty{\sim80}{\percent}\ (21.75<F814W<22.9$ mag). Main plot shows full population of UCD candidates. Counts in each bin are included and density contour scaling is the same for each plot for comparative purposes. Although minimal in number, the brightest UCDs are clearly associated with the elliptical NGC 4889 (bin 1) but with agglomerations notable around NGC 4874 and IC 4051. The smaller magnitude range of bin 2 encompasses a larger and more spatially distributed population of dimmer UCDs. This suggests that intracluster UCDs are less likely to be bright (\ie \qty{>21.75}{\mag}), as the majority of the intracluster UCDs are to be found in bin 2. North is up and East is left.}
    \label{fig:Coma_UCDs_size_map}
\end{figure*}

In exploring intracluster GCs and UCDs within the core (\qty{\sim1}{\mega\parsec}) of Coma, we find that only 14\% of UCDs are to be found outside $8\times R_e$, compared to 24\% of GCs. Averaging over the 196 member galaxies in our sample, we demonstrate that the GC and UCD populations decline radially following a S\'{e}rsic profile with index $n\sim1$.

\subsection{Distribution of UCDs in Coma by luminosity}\label{sec:UCD_sizes}
 In \cref{fig:Coma_UCDs_size_map} we show a number of views of the spatial distribution of UCD candidates in our data, with a density map overlaid on a background Coma cluster image to provide context. From the splits of data detailed below, a 2D-histogram of the object positions was created on a $30\times21$ grid (to maintain the WCS aspect ratio), to which we then apply Gaussian smoothing. We use \texttt{matplotlib.contourf} overlaid on the SDSS g-band image to produce a density contour map. The UCD candidate positions are also included. Using the mag F814W data for the UCDs, we split into 2 magnitude bins, $16.0<F814W<21.75$ mag and $21.75<F814W<22.9$ mag, respectively, to visualize the distribution of brightest and dimmest candidates, finding an $\sim80:20$ split by magnitude ($\sim20:80$ by count) to be informative. The count of UCDs in each bin is included for reference and the density colorbar scaling has been adjusted to correlate in each plot for comparison. 
 
 Three density peaks of luminosity are evident in \cref{fig:Coma_UCDs_size_map}, in both the bright magnitude bin ($\qty{\sim20}{\percent}$ by total - bin 1) and also the more populous, but dimmer magnitude bin ($\qty{\sim80}{\percent}$ by total - bin 2). These peaks are visually congruent with the galaxy cores of the three main ellipticals, IC 4051, NGC 4889 and NGC 4874, as noted in §\ref{sec:Spatial_dist}. However, we confirm this by executing a procedure similar to that carried out in §\ref{sec:Radial_dist}, where we estimate the candidate UCD population within $8R_e$ of a host galaxy. Here though, we limit the sample to the `bright' magnitude range, $16.0\!<\!F814W\!<\!21.75$ mag, containing 105 UCDs, and the dim magnitude range $21.75\!<\!F814W\!<\!22.9$ mag, containing 418 UCDs. The counts of UCDs within $8R_e$ of the host galaxy center for the top 8 galaxies, are shown in Table \ref{tab:CSS_Count_mag_bins}.
 
\begin{deluxetable}{cccccc}
\centering
\label{tab:CSS_Count_mag_bins}
\tablecaption{Count of bright ($T_b=105$) and dim ($T_d=418$) UCDs within $8R_e$ of a host galaxy, illustrating the dominant nature of the three giant ellipticals hosting UCD candidates. Uncertainties on counts are $1\sigma$ Poisson confidence intervals.}
\tablehead{
	\multirow{2}{*}{Galaxy} & \multicolumn{2}{c}{Bright}  & \multicolumn{2}{c}{Dim} & Total \\ 
	& \colhead{$N_b(\pm)$} & \colhead{$N_b/T_b$} & \colhead{$N_d(\pm)$} & \colhead{$N_d/T_d$} & \colhead{$N_b+N_d$}
}
\startdata
	NGC 4889 & 40(6) & 0.38 & 95(10) & 0.23 & 135 \\ 
	NGC 4874 & 19(4) & 0.18 & 101(10) & 0.24 & 120 \\ 
	IC 4051 & 12(3) & 0.11 & 61(8) & 0.15 & 73 \\ 
	NGC 4882 & 5(2) & 0.05 & 21(5) & 0.05 & 26 \\ 
	NGC 4873 & 5(2) & 0.05 & 20(4) & 0.05 & 25 \\ 
	NGC 4871 & 3(2) & 0.03 & 20(4) & 0.05 & 23 \\ 
	IC 4011 & 4(2) & 0.04 & 10(3) & 0.02 & 14 \\ 
	IC 3998 & 2(1) & 0.02 & 10(3) & 0.02 & 12 
\enddata
\end{deluxetable}
 The densest agglomeration of brightest UCDs is visually associated with the binary BCG elliptical, NGC 4889, and this is confirmed in Table \ref{tab:CSS_Count_mag_bins} with $\sim\!38\%$ of the total bright bin being found within $8R_e$ of NGC 4889. However, as would be expected from the GCLF, the majority of the UCDs are dimmer, as shown in bin 2. With the exception of the three main clumps, there are no other significant density groupings visible, \ie $N_\text{UCD}>0.1N_\text{Total}$ within $8R_e$. We note that the emergence of the intracluster UCDs is also clear. This observation is consistent with the result of the intracluster UCD analysis (§\ref{sec:ICUCDs}), insomuch as UCDs have a higher probability of being located close to a host galaxy core than the dimmer GCs. The wider extent and dispersion of the agglomeration around NGC 4874, compared to that around NGC 4889, is also of note, despite the similar numbers of UCDs out to \qty{100}{\kilo\parsec} (see \cref{tab:CSS_radial_dist}). This wider dispersion is clearly illustrated in the comparison panels of \cref{fig:Coma_UCDs_Location_2pcf_Grid} for NGC 4889 and NGC 4874.

 \cref{fig:Coma_UCDs_size_map} clearly shows clustering of UCDs congruent with the centers of NGC 4889 and IC 4051. However, the dispersion of UCDs around NGC 4874, and low fraction of bright UCDs for this giant elliptical compared to the fraction of dimmer UCDs, which increase above that seen around NGC 4889, as noted in Table \ref{tab:CSS_Count_mag_bins}, is suggestive of
 significant past merger events and a level of merging activity commensurate with the dispersion of the central UCDs.

\begin{figure*}
    \centering
    \includegraphics[width=0.8\linewidth]{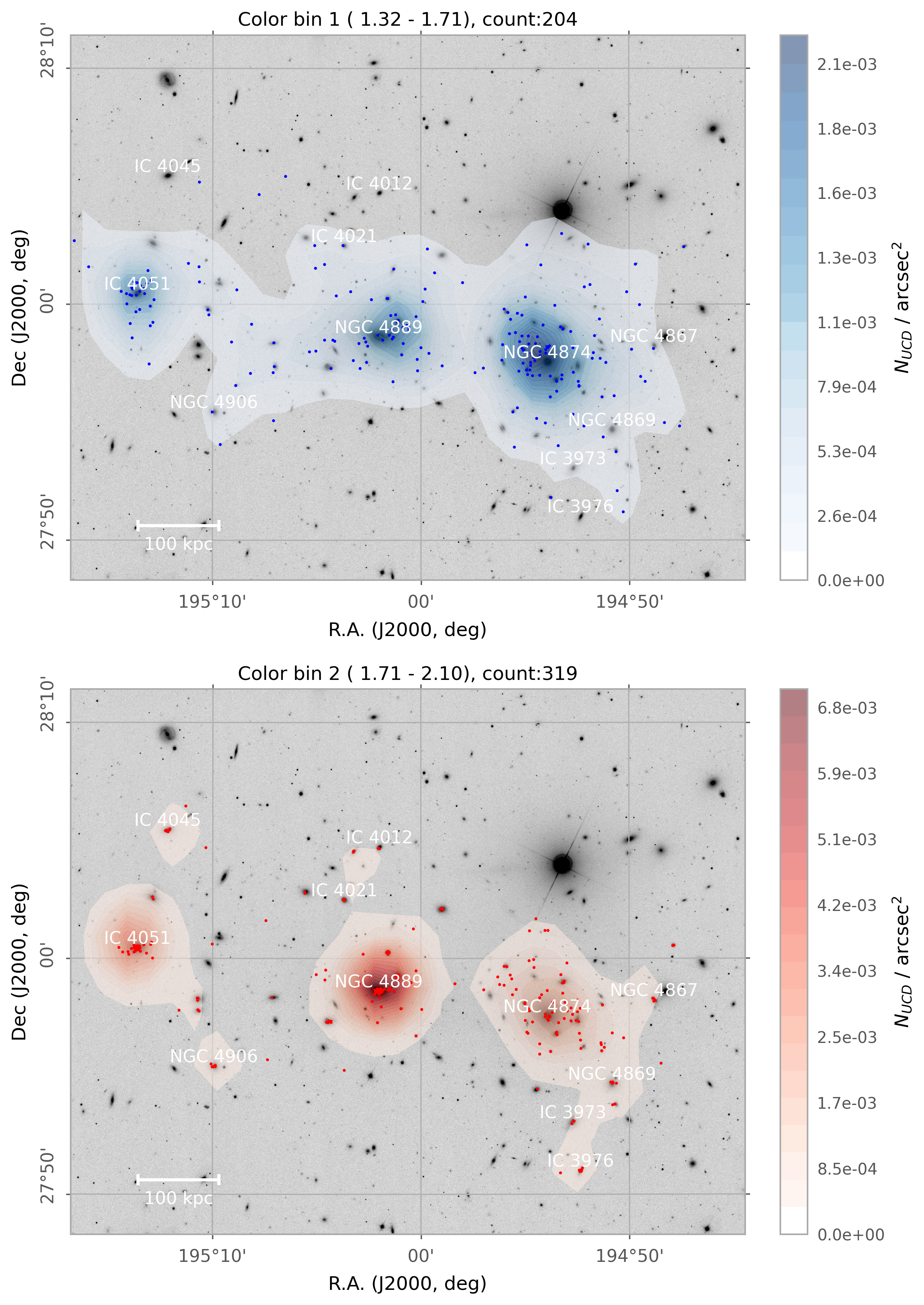}
    \caption{Spatial distribution of UCD candidates from our data, split into nominal red and blue bins by bimodal GMM intersection color for the UCD magnitude ranges from Table \ref{tab:cmd_unconstrained_fits},
    as described in the text. Density contours are included. Upper panel shows blue sample of UCD candidate population ($1.32\leq(F475W - F814W)\lesssim1.71$) and the lower panel shows red sample of UCD candidate population ($1.71\lesssim(F475W - F814W)\leq2.10$). The blue UCD candidates show a greater dispersion about the three giant ellipticals than exhibited by the red UCD candidates. Conversely, the red UCDs exhibit greater density and clustering about the centers of many of the galaxies, in addition to the three main ellipticals. North is up and East is left.}
    \label{fig:Coma_UCDs_color_map}
\end{figure*}

\subsection{Distribution of UCDs in Coma by color}\label{sec:UCD_color_split}
\begin{figure}[ht]
    \centering
    \includegraphics[width=1\linewidth]{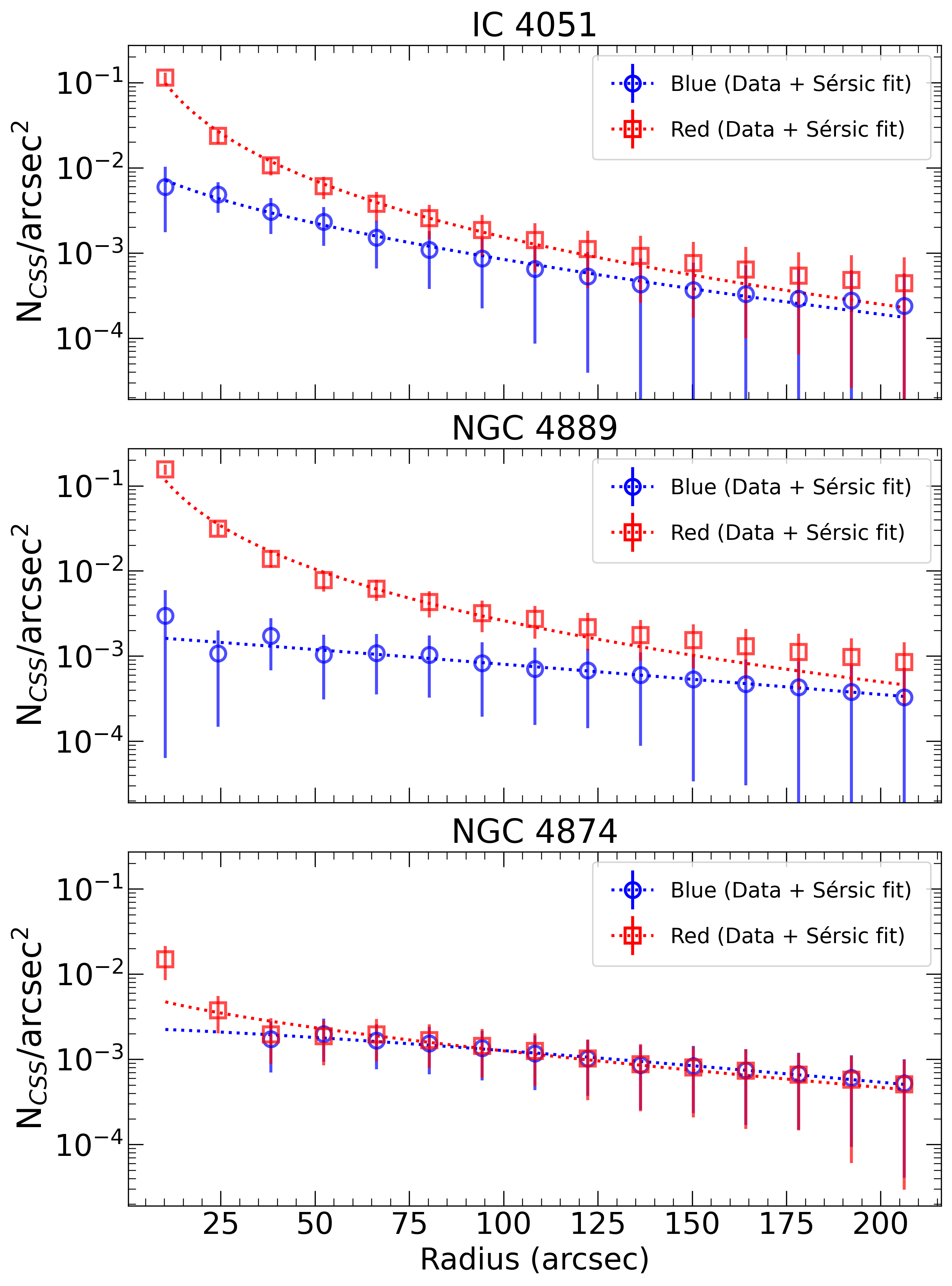}
    \caption{Radial color profile of UCDs about the three giant ellipticals. The color split used correlates with that shown in \cref{fig:Coma_UCDs_color_map} and Table \ref{tab:CSS_Count_col_bins}, as described in the text. The higher central density of red UCDs is clear. The comparable density of red and blue UCDs for NGC 4874 outside \qty{\sim35}{\arcsec} (\qty{\sim17}{\kilo\parsec}) is also of note.}
    \label{fig:UCD_color_radial_profile}
\end{figure}

In \cref{fig:Coma_UCDs_color_map}, we present the distribution of UCDs by color. The blue and red sequences of UCDs are defined at the intersection of the two Gaussians that fit the bimodal population as listed in the UCD relevant magnitude range GMMs from Table \ref{tab:cmd_unconstrained_fits}. For the `bright' magnitude range ($16.0<F814W<=22.60$) a blue-red color threshold of $(F475W - F814W)=1.71$ was used, and for dimmer UCDs ($F814W>22.61$) the split was set at $(F475W - F814W)=1.63$. The same split of color was used in all subsequent analysis, including the estimated UCD populations, within $8R_e$ of a galaxy host, as shown in Table \ref{tab:CSS_Count_col_bins}, and also the radial color profiles about the three giant ellipticals, IC 4051, NGC 4889 and NGC 4874, as shown in \cref{fig:UCD_color_radial_profile}. From this blue-red split of UCD candidates a 2D-histogram of the posittion of UCD candidates was created with the same methods as \cref{fig:Coma_UCDs_size_map}. We use \texttt{matplotlib.contourf} overlaid on the SDSS g-band image to produce a density contour map. The UCD candidate positions are included.

\begin{deluxetable}{cccccc}
\centering
\label{tab:CSS_Count_col_bins}
\tablecaption{Count of blue ($T_b=204$) and red ($T_r=319$) UCDs within $8R_e$ of a host galaxy. Uncertainties on counts are $1\sigma$ Poisson confidence intervals. The results illustrate the greater density of red UCDs within $8R_e$ of the three giant ellipticals, as shown in \cref{fig:Coma_UCDs_color_map}.}
\tablehead{
	\multirow{2}{*}{Galaxy} & \multicolumn{2}{c}{Blue}  & \multicolumn{2}{c}{Red} \\ 
	& \colhead{$N_b(\pm)$} & \colhead{$N_b/T_b$} & \colhead{$N_r(\pm)$} & \colhead{$N_r/T_r$} & \colhead{Total}
}
\startdata
	NGC 4889 & 32(6) & 0.16 & 103(10) & 0.32 & 135 \\ 
	NGC 4874 & 61(8) & 0.30 & 59(8) & 0.18 & 120 \\ 
	IC 4051 & 21(5) & 0.10 & 52(7) & 0.16 & 73 \\ 
	NGC 4882 & 9(3) & 0.04 & 17(4) & 0.05 & 26 \\ 
	NGC 4873 & 12(3) & 0.06 & 13(4) & 0.04 & 25 \\ 
	NGC 4871 & 9(3) & 0.04 & 14(4) & 0.04 & 23 \\ 
	IC 4011 & 3(2) & 0.01 & 11(3) & 0.03 & 14 \\ 
	IC 3998 & 5(2) & 0.02 & 7(3) & 0.02 & 12 
\enddata
\end{deluxetable}

The contours and spatial distribution of the candidate UCDs in \cref{fig:Coma_UCDs_color_map}, clearly show that the blue (metal-poor) UCDs are more dispersed than the red (metal-rich) UCDs around the three giant ellipticals in the region of the Coma cluster under analysis. The red UCD candidates exhibit a higher density in the central regions of the three giant ellipticals, NGC 4889, NGC 4874 and IC 4051, than the blue UCDs. Of note, is the complete lack of blue UCDs co-spatial with the core of NGC 4874, as confirmed in the color radial profile \cref{fig:UCD_color_radial_profile}, where with same color samples are used. We note that this is despite NGC 4874 having nearly a third ($29\%$) of the blue UCDs within $8R_e$ with $19\%$ of the red UCDs, as detailed in Table \ref{tab:CSS_Count_col_bins}. Conversely, NGC 4889 has only $15\%$ of the blue UCDs within $8R_e$ compared to $32\%$ of the red UCDs. As illustrated in \cref{fig:UCD_color_radial_profile} the radial density of the red UCDs is an order of magnitude greater than the blue UCDs about IC 4051 for a projected radius of \qty{\sim25}{\arcsec} (\ie \qty{\sim12}{\kilo\parsec}). For NGC 4889 this order of magnitude disparity is maintained out to at least \qty{\sim50}{\arcsec} (\qty{\sim25}{\kilo\parsec}). The red UCD candidates are also seen to be associated with the centers of other galaxies in this region of Coma, with two other islands of density around IC 4045, NGC 4906 and bridges from the dispersion around NGC 4874, linking NGC 4867, NGC 4869, IC 3973 and IC 3976, as illustrated in \cref{fig:Coma_UCDs_color_map}. The existence of red UCDs with higher luminosities, especially in the central region of NGC 4889, can be explained by the general color trend of the CMD (see \cref{fig:Coma_CSS_CMD} and discussed in §\ref{sec:CMD_GCs_and_UCDs}), insomuch as the more luminous UCDs are redder. As discussed in the previous section (§\ref{sec:Spatial_dist}), the red UCDs are more likely to be found closer to a host galaxy. The structure noted in §\ref{sec:UCD_sizes} around NGC 4874 is predominantly associated with bluer UCDs. Conversely, the structure congruent with the nucleus of NGC 4889 is also observed in the split of red UCD candidates.

\section{Conclusions} \label{sec:Conclusions}
Using data from the HST/ACS Coma Cluster Survey we have identified \RTPVarUCDTotal UCD candidates in \RTPVarCSSTotal CSSs. The UCDs have been shown to exhibit a mass-metallicity relation (blue-tilt) consistent with literature color trend models. 

We also demonstrate the departure of the sample from a simple Gaussian fit to the GCLF, with an excess at the bright end, further confirming the presence of UCDs in our sample and multiple formation pathways for the composite population of objects within this parameter space. From the excess of the population of UCDs to a GCLF model, we estimate a lower limit for the number of UCDs to have formed through a process other than growth as massive GCs in the Coma cluster as $N_\text{UCD}\!\gtrsim\!32\!\pm\!1$. Furthermore, we estimate $\approx252\pm6$ or $66\%$ of UCDs with a cluster mass $\Mstar\gtrsim10^7 \Msun$ have formed through a process distinct from GC growth. We also estimate a total predicted count for CSS in the surveyed regions of the Coma cluster to be $N_{CSS}\!\approx\!\num[group-separator = {,}]{69400(1400)}$.

By analysis of the radial distribution of UCDs, we demonstrate that, by ratio, UCDs have a higher probability of being found closer to one of the three giant ellipticals, NGC 4889, NGC 4874 and IC 4051. We find $34\pm4.4\%$ of UCDs within a projected radial distance of \qty{25}{\kilo\parsec} of one of these three ellipticals, compared to $22\pm0.5\%$ of GCs. The dominance of UCDs, by ratio, with radial distance from the giant ellipticals continues out to \qty{100}{\kilo\parsec}. The radial distribution of UCDs in the central \qty{1}{\mega\parsec} of Coma cements the status of IC 4051 in comparison to the binary BCGs (NGC 4889 and NGC 4874) in that the populations of UCDs about these three galaxies are of similar order.

The spatial distribution of UCDs in comparison to effective radii for all the galaxies in the central part of the cluster also demonstrate that while the majority of the brightest UCDs are grouped around NGC 4889 $38\pm5.7\%$, NGC 4874 and IC 4051 are also notable, hosting bright UCD populations of $19\pm3.8\%$ and $11\pm3.8\%$ respectively. We also confirm that, in general, UCDs are more likely to be located closer to a host galaxy, with the ICUCD fraction being $\sim\!14\%$ compared to an ICGC of $\sim\!24\%$ at a similar host effective radius of $R\simeq8R_e$.

Although we have shown conglomerations of UCDs around three of the main galaxies in the central \qty{1}{\mega\parsec} of Coma, as discussed in §\ref{sec:Radial_dist} and shown in \cref{fig:Coma_UCDs_Location_2pcf_Grid}, we observe a greater radial dispersion of the UCD population around the binary BCG NGC 4874, in addition to a minimal number of blue central UCDs, suggesting NGC 4874 has been through more recent episodes of merging activity.

We find similar conglomerations around the three giant ellipticals in our color analysis of the UCD candidate population, as discussed in §\ref{sec:UCD_color_split} and shown in \cref{fig:Coma_UCDs_color_map,fig:UCD_color_radial_profile}, showing that red (metal-rich) UCDs are more closely associated with the central regions of galaxies. We also demonstrate that the red UCDs have a higher surface density, with blue (metal-poor) UCDs showing lower concentration and wider distribution in comparison. 

A NIRCAM general observation proposal for JWST observation of the Coma cluster will image more than 100 fields centered on 39 of the largest elliptical galaxies in Coma using F150W and F365W bands \citep[Cycle 3, ID. \#5989][]{Jensen_2024}. Planned for mid-2025 the program will focus on measuring surface brightness fluctuations (SBF) in the cluster, the calibration of which will enable an independent, high-precision determination of the cosmological distance scale in the local universe. However, parallel observations with the Near-Infrared Imager and Slitless Spectrograph (NIRISS) will supplement the dataset available for the surrounding Coma GC and UCD populations. Thus, this survey will likely provide the data to confirm the nature of a significant fraction of the UCD candidates presented here and will yield interesting results regarding the chemical composition of UCDs across Coma. 

Although not planned in the above proposal, with sufficiently sensitive future observations of velocity dispersion, mass estimation of observed UCDs would be possible and thus identification of excess dynamical mass compared to a canonical stellar population. Consequently, nuclear IMBH or SMBH mass could be estimated, providing further evidence to answer the outstanding question of UCD formation processes.

\section*{Acknowledgments}
We thank the anonymous referee for a prompt and constructive report that helped us to improve this paper.

We thank Blanca O. Garcia and Katrina Martinez, at The University of Texas Rio Grande Valley, for their help and support for this project.

Based on observations made with the NASA/ESA Hubble Space Telescope, obtained at the Space Telescope Science Institute, which is operated by the Association of Universities for Research in Astronomy, Inc., under NASA contract NAS5-26555. These observations are associated with programs GO 10861, 11711, 12918. 

This research has made use of the NASA Astrophysics Data System Bibliographic services (ADS), funded by NASA under Cooperative Agreement 80NSS-C21M00561. 

This research has made use of the NASA/IPAC Extragalactic Database (NED), which is funded by the National Aeronautics and Space Administration and operated by the California Institute of Technology.

This work is supported by the National Science Foundation under Cooperative Agreement PHY-2019786 (The NSF AI Institute for Artificial Intelligence and Fundamental Interactions, \url{http://iaifi.org/}).

\bibliography{references}{}
\bibliographystyle{aasjournal}



\end{document}